\documentclass[pra,twocolumn,superscriptaddress,showpacs,nobibnotes,notitlepage,nofootinbib]{revtex4-2}

\usepackage[utf8]{inputenc}
\usepackage[margin=1in]{geometry} 
\usepackage{amsmath,amsthm,amssymb,hyperref}
\hypersetup{
    colorlinks=true,
    linkcolor=blue,
    citecolor=magenta,
    filecolor=magenta,      
    urlcolor=blue,
    pdftitle={},
    pdfpagemode=FullScreen,
    }
\usepackage{adjustbox}
\usepackage{graphicx}
\usepackage{caption}
\usepackage{float}
\usepackage{subcaption}
\usepackage{xcolor}
\usepackage{soul}
\newtheorem{theorem}{Theorem}
\begin{document}

\title{Qubit recycling and the path counting problem}
\author{Zijian Song}
\affiliation{Department of Physics, University of California, Davis, CA 95616, USA}
\author{Isaac H. Kim}
\affiliation{Department of Computer Science, University of California, Davis, CA 95616, USA}
\date{\today}

\begin{abstract}
Recently, it was shown that the qudits used in circuits of a convolutional form (e.g., Matrix Product State sand Multi-scale Entanglement Renormalization Ansatz) can be reset unitarily \href{https://doi.org/10.1103/PhysRevA.103.042613}{[Phys. Rev. A 103, 042613 (2021)]}, even without measurement. We analyze the fidelity of this protocol for a family of quantum circuits that interpolates between such circuits and local quantum circuits, averaged over Haar-random gates. We establish a connection between this problem and a counting of directed paths on a graph, which is determined by the shape of the quantum circuit. This connection leads to an exact expression for the fidelity of the protocol for the entire family that interpolates between convolutional circuit and random quantum circuit. For convolutional circuits of constant window size, the rate of convergence to unit fidelity is shown to be $\frac{q^2}{q^2+1}$, independent of the window size, where $q$ is the local qudit dimension. Since most applications of convolutional circuits use constant-sized windows, our result suggests that the unitary reset protocol will likely work well in such a regime. We also derive two extra results in the convolutional limit, which may be of an independent interest. First, we derive exact expressions for the correlations between reset qudits and show that they decay exponentially in the distance. Second, we derive an expression for the the fidelity in the presence of noise, expressed in terms of the quantities that define the property of the channel, such as the entanglement fidelity. 
\end{abstract}

\maketitle

\section{Introduction}
\label{sec:intro}

In recent years, there has been a tremendous amount of progress in quantum computing technology. Demonstrations of quantum computational supremacy~\cite{Arute2019,Zhong2020,Madsen2022} indicates that we are entering an era in which the quantum computers that are available today can perform computational tasks that are likely difficult for the existing classical computers. An outstanding open problem at this point is whether there are practical problems of interest that can be solved using these computers.

One approach that garnered a lot of attentions in recent years is the variational quantum eigensolver (VQE)~\cite{Peruzzo2014,Farhi2014}. This is an approach in which one views a quantum computer as a device capable of preparing some quantum state. Then the variational principle dictates that, given a Hamiltonian, the energy obtained from this quantum state will upper bound the true ground state energy. Thus, the main idea behind this approach is to judiciously optimize the variational energy obtained from a quantum computer, thereby hoping to obtain an accurate enough approximation to the ground state energy. 

Early instantiations of VQE involved quantum circuits that are targeted towards approximating ground states of chemical systems~\cite{Peruzzo2014,Kandala2017}. More recently, ansatzes that have been studied in the context of simple models of locally interacting quantum many-body systems~\cite{White1992,Verstraete2004,Vidal2008} were adopted. Examples of such ansatzes include holographic quantum circuits~\cite{Kim2017,Kim2017a,Foss-Feig2021,Chertov2021} and deep entanglement renormalization ansatz~\cite{Kim2017b}. These ansatzes have several notable advantages, including its intrinsic noise-resilience, benign requirement on the number of qubit, and theoretical evidence that these ansatzes can approximate physical ground states of interest with a moderate number of qubits and circuit size.

One common thread that pierces through all these approaches is the underlying structure of the circuit; they are all in the convolutional form~\cite{Poulin2009}. Convolutional circuit is a circuit in which there is a sliding active window of circuits that act nontrivially on a subset of qubits. It was shown in Ref.~\cite{Anikeeva2021} that, by exploiting the convolutional structure, it is possible to reset some of the qubits unitarily, even without using any physical reset operations. Such reset operations can be useful because they can dramatically reduce the number of qubits needed to implement the convolutional circuit on a quantum computer.

\begin{figure}[h]
    \centering
    \begin{subfigure}[b]{0.45\textwidth}
         \centering
         \includegraphics[height=3cm]{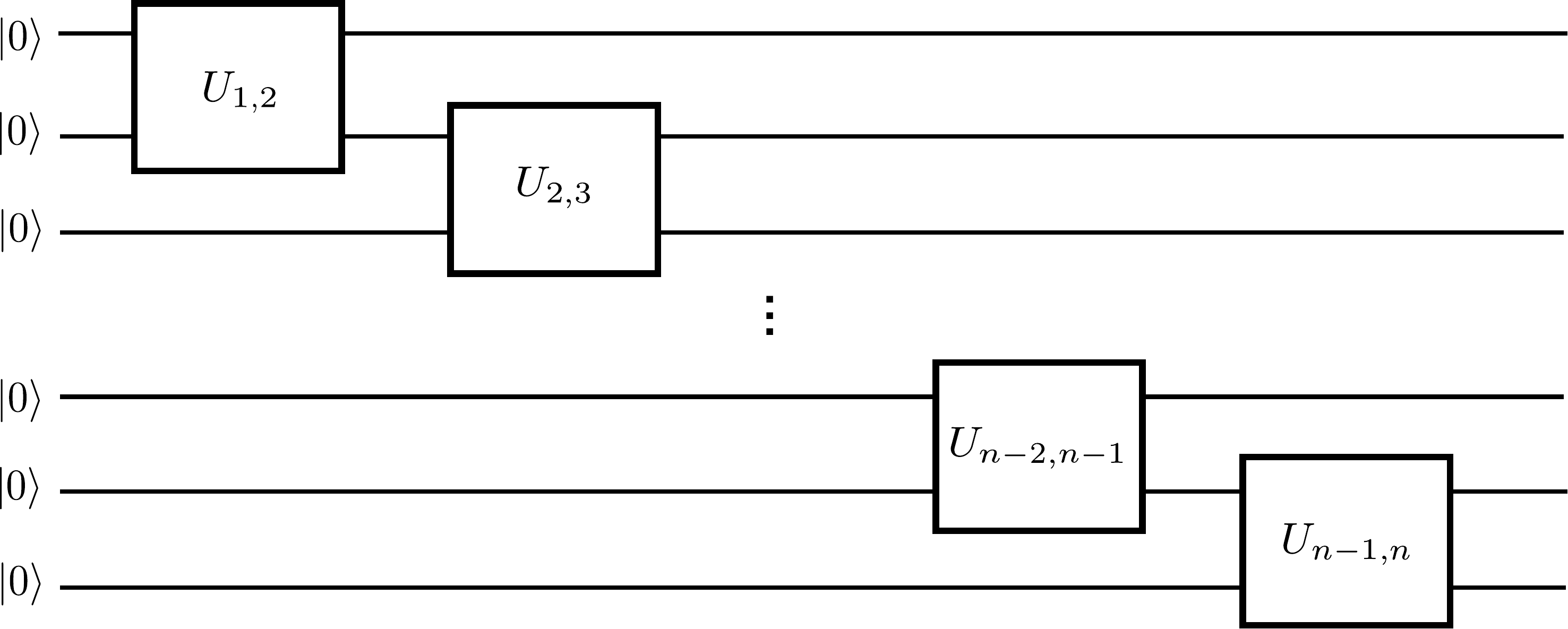}
         \caption{}
         \label{rewind_convolutional}
     \end{subfigure}
     \hfill
    \begin{subfigure}[b]{0.45\textwidth}
         \centering
         \includegraphics[height=3cm]{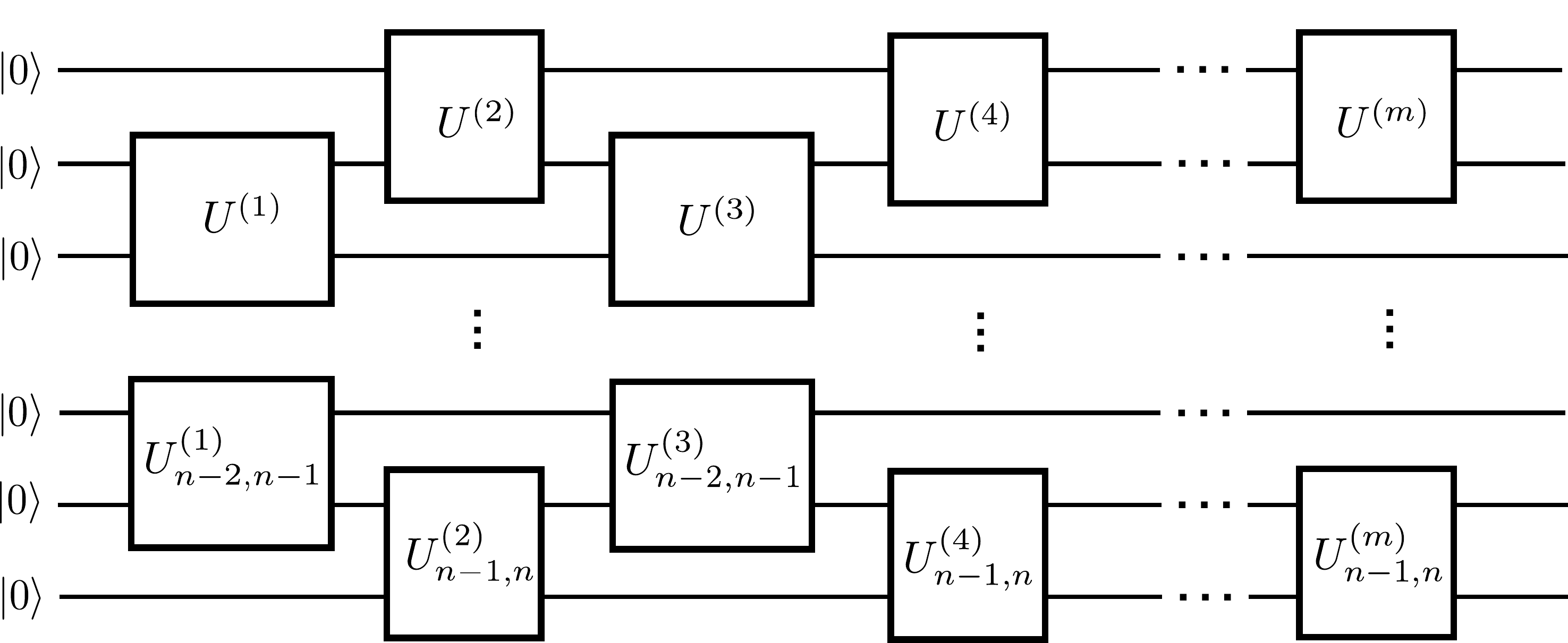}
         \caption{}
         \label{rewind_local}
     \end{subfigure}
     \hfill
    \begin{subfigure}[b]{0.45\textwidth}
         \centering
         \includegraphics[height=3.1cm]{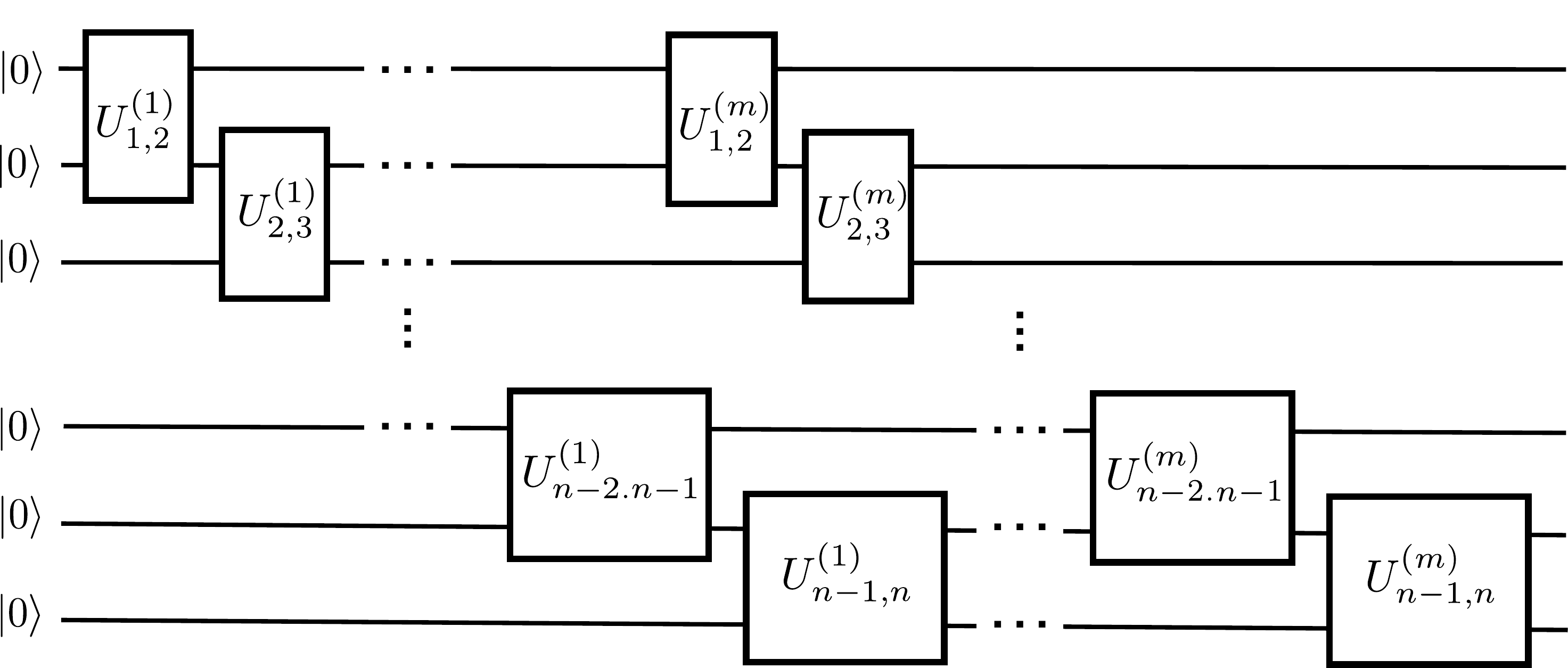}
         \caption{}
         \label{rewind_hybrid}
     \end{subfigure}
     \caption{An interpolation between convolutional circuit and random quantum circuit. (a) Convolutional limit (b) Random quantum circuit limit. The blue blocks are random unitaries. (c) A middle ground between (a) and (b). The two limits can be interpolated by changing the window size. }
     \label{fig:interpolation}
\end{figure}

However, precisely how many qubits one can save using this approach depends on a certain data about the circuit which is difficult to estimate. Mathematically speaking, Ref.~\cite{Anikeeva2021} showed that the eigenvalue gap of a certain matrix constructed from the convolutional circuit determines the rate at which the underlying protocol converges; the larger the gap, the better the protocol works. However, it is not clear at all what the value of the gap will be. In particular, in practical applications of convolutional circuits, one would often consider a family of circuits in which the size of the sliding window increases; see FIG.~\ref{fig:interpolation} (c) for example. A natural question then is how well the protocol work as the window size increases. Does the gap decrease with the window size or not? If it does, how does it scale?

While answering these questions analytically for an arbitrary quantum circuit is hopeless, we can at least hope to understand this question for the \emph{typical instances}, by taking an average over randomly chosen gates. Taking average over randomly chosen gates is a well-established technique in quantum information theory, which we utilize in this paper. We show that there is a universal decay rate of $\frac{q^2}{q^2+1}$ if the window size is constant, where $q$ is the local qudit dimension. This means that, even for a realistic family of convolutional circuits one expects to be used in variational calculations, we can expect the protocol in Ref.~\cite{Anikeeva2021} to continue to work well.

In the convolutional limit (see FIG.~\ref{fig:interpolation}.), we establish two extra facts. First, we show that the correlation decays exponentially. Second, we derive an expression for the fidelity in the presence of arbitrary noise. This expression is applicable to any noise channel, and conveniently expressed in terms of the quantities that define the property of the channel, such as the entanglement fidelity~\cite{Schumacher1996}. 

Our derivations rest on the idea that the second moment of the a quantum circuit with Haar-random gates can be mapped to a statistical mechanics models~\cite{Nahum2018,Hunter-Jones2019}. This mapping works quite flexibly, independent of the shape of the circuit and boundary conditions, a fact that we utilize in this paper. Moreover, for the cases in which noise is present, the mapping works independent of the details of the noise.

The rest of this paper is structured as follows. In Section~\ref{sec:rewinding} we briefly review the rewinding protocol in Ref.~\cite{Anikeeva2021}. In Section~\ref{sec:circuits_to_lattice}, we explain how our model is mapped to a statistical mechanics models. Parts of this discussion are based on Ref.~\cite{Nahum2018,Hunter-Jones2019} but the material about path-counting is new to the best of our knowledge. In Section~\ref{sec:recycling_perfect}, leveraging the path counting formulae, we derive exact expressions for the fidelity of the rewinding protocol under various circumstances. In Section~\ref{sec:recycling_imperfect}, we explain how to incorporate the effect of noise and explain how our calculation in the convolutional limit can be modified. We conclude with a set of open problems in Section~\ref{sec:conclusion}.

\section{Rewinding Protocol}
\label{sec:rewinding}

In this Section, we briefly review the rewinding protocol in Ref.~\cite{Anikeeva2021} and set up the notation for the rest of the paper. 

Consider a sequence of quantum gates applied to the initial state $|0^n\rangle$, where $n$ is the number of qudits. The sequence can be written as
\begin{align}
    \mathcal{C} = (g_1, g_2, ..., g_{N-1}, g_{N}).
\end{align}
where $g_i$ represents the $i$'th gate. This sequence implements a unitary
\begin{align}
    U(\mathcal{C}) := g_N g_{N-1} ... g_1.
\end{align}

\begin{figure}[h]
    \centering
    \includegraphics[width = 7cm]{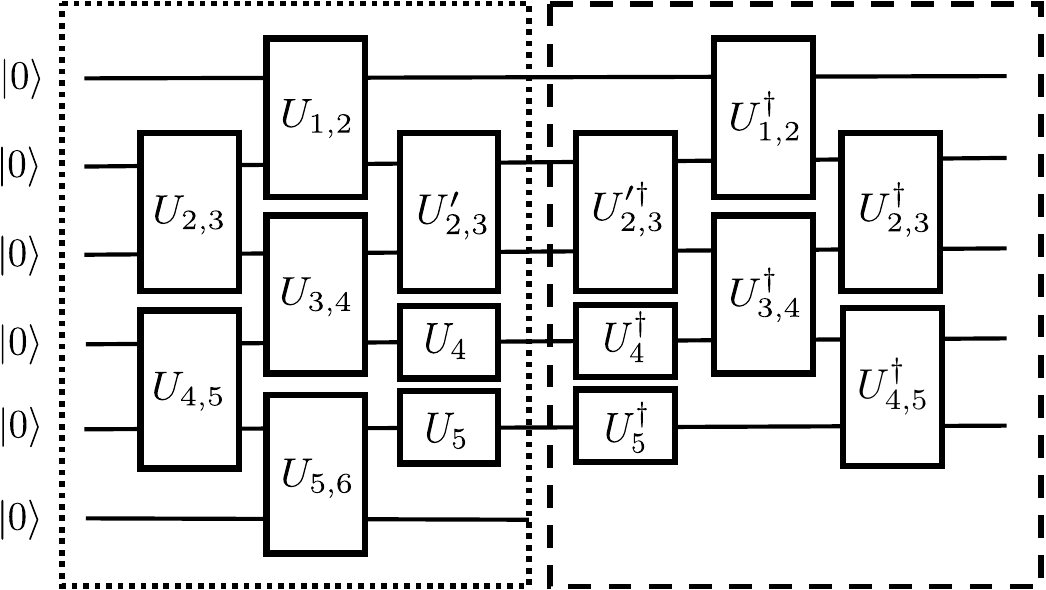}
    \caption{An example of the rewinding circuit. The first five qudits are idle. And the subcircuit $\mathcal{R}(\mathcal{C}_{I,1})$ in the dashed box is the rewinding circuit of the circuit $U(\mathcal{C})$ in the dotted box.}
    \label{rewinding}
\end{figure}

In the rewinding protocol, we are interested in reusing some of qudits that no longer participate in the computation at some point for other purposes. Without loss of generality, decompose the unitary $U(\mathcal{C})$ as
\begin{align}
    U(\mathcal{C}) = U(\mathcal{C}_2) U(\mathcal{C}_1),
\end{align}
in which
\begin{align}
    \mathcal{C}_1 &= (g_1, ... , g_k), \notag \\
    \mathcal{C}_2 &= (g_{k+1}, ..., g_{N}),
\end{align}
wherein we assume there is a subset of qudits on which $\mathcal{C}_2$ acts trivially. We refer to those qudits as \emph{idle} qudits. The main goal is to convert some of the idle qudits to $|0\ldots 0\rangle$. Consider a subset of $\mathcal{C}_1$ consisting of gates that only act on the idle qudits. 
\begin{align}
    \mathcal{C}_{I,1} := (\tilde{g}_1, ..., \tilde{g}_{N'})
\end{align}
The rewinding circuit is thus defined as,
\begin{align}
    \mathcal{R}(\mathcal{C}_{I,1}) := (\tilde{g}^{\dagger}_{N'}, ..., \tilde{g}^{\dagger}_{1}).
\end{align}
After we apply the rewinding circuit, the state becomes
\begin{align}
    |\psi_n\rangle := \mathcal{R}(\mathcal{C}_{I,1}) U(\mathcal{C}) |0^n\rangle. \label{eq:rewinding}
\end{align}
An illustration of the rewinding circuit is shown in Fig.~\ref{rewinding}.

In order to assess the performance of the rewinding protocol, we use the \emph{fidelity} as a figure of merit. Without loss of generality, suppose we are given a circuit $\mathcal{C}$ and then applied a rewinding protocol, applied to an initial state $|0\ldots 0\rangle$ over $n$ qudits. Denoting this state as $|\psi_n\rangle$, our goal is to compute the fidelity, defined as
\begin{align}
    F = \langle \psi_n | (|0\rangle \langle 0 |_1 \otimes I) | \psi_n\rangle, \label{eq:fidelity}
\end{align}
where $|0\rangle \langle 0 |_1$ is the projector onto the $|0\rangle$ state for the first qudit and $I$ is the identity operator on the remaining $n-1$ qudits.

In fact, instead of computing Eq.~\eqref{eq:fidelity} for a single circuit $\mathcal{C}$, we will be averaging Eq.~\eqref{eq:fidelity} over some ensemble of unitaries:
\begin{align}
    F^{avg} = \int F\ d\mu(\{U_i\}). \label{eq:avg_fidelity}
\end{align}
The averaging serves two purposes. First and foremost, this often leads to analytic expressions, which are useful to understand the performance of the protocol. Second, averaging lets us understand how well the protocol works well for typical circuits.

In Ref.~\cite{Anikeeva2021} the averaged fidelity was estimated via numerical means, against a \emph{convolutional circuit}. A convolutional circuit is a circuit in which a fixed-size active window moves over the qudits of the circuit. The general form can be written as
\begin{align}
     U_{[n-k+1,n]} \cdot \cdot \cdot U_{[2,k+1]} U_{[1,k]},
     \label{eq:convolutional_def}
\end{align}
where $U_{[i,j]}$ is the unitary that acts from the $i$-th qudit to the $j$-th qudit; see Fig.~\ref{fig:interpolation}(a) for an example. 

In this paper, we generalize the study in Ref.~\cite{Anikeeva2021} substantially through a new analytic technique developed in this paper. This technique is applicable to more general family of quantum circuits that include convolutional circuit as a special case; instead of demanding that the individual gates in Eq.~\eqref{eq:convolutional_def} are two-qudit gates, we can replace them to a finite-depth quantum circuit acting on a constant-sized window. In the limit the window size becomes large, the resulting circuit can be viewed as a local quantum circuit.






\section{Circuits and path counting problem}
\label{sec:circuits_to_lattice}

Our approach to computing Eq.~\eqref{eq:avg_fidelity} is to establish a connection between the averaged fidelity and a \emph{path counting problem} on a square lattice. Our key observation is that the averaged fiedelity can be expressed as a summation over the number of directed paths satisfying some constraints, weighted by appropriate factors that only depend on $q$. This observation is useful because the number of directed paths often have an exact expression, whose solutions can be found, for instance in Ref.~\cite{krattenthaler2015lattice}.

To establish this connection, we shall proceed in the following steps. First, we shall use the approach in Ref.~\cite{Nahum2018,Hunter-Jones2019} to map the fidelity to a partition function of some statistical mechanics model. It was shown in these references that the resulting partition function can be expressed as a sum of contributions, each of which come from the lengths of the domain wall. While an arbitrary number of domain walls is allowed in these setups, for us, it will suffice to only consider a \emph{single} domain wall. This will be our main source of simplification. We can attribute this fact to a boundary condition that arises in our setup, which we explain in Section~\ref{sec:Mapping}.

\begin{figure*}
    \centering
    \includegraphics[width=16cm]{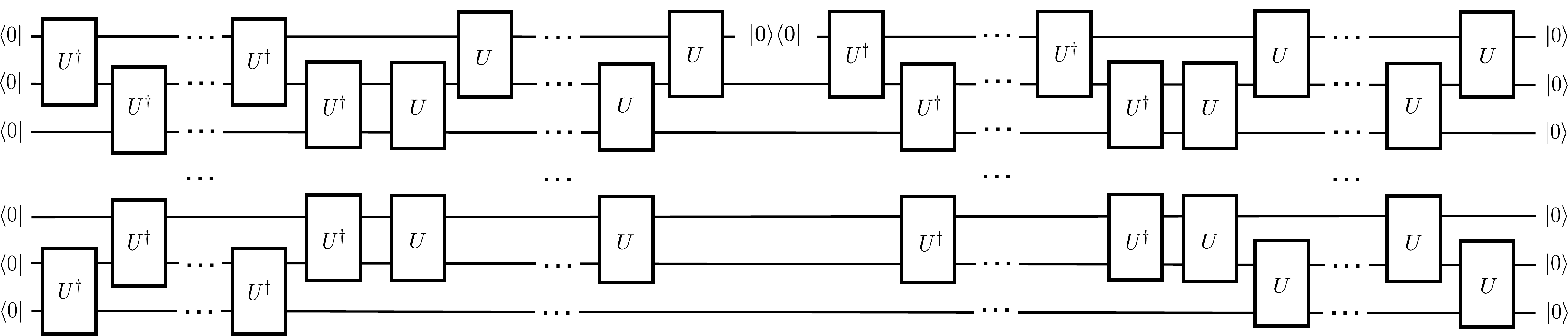}
    \caption{A diagrammatic representation of the fidelity of recycling the first qudit after we apply the rewinding protocol. Here $U$ are the 2-site Haar random unitaries.}
    \label{fig:local_circuit}
\end{figure*}

The rest of this Section is organized as follows. Section~\ref{sec:folding},~\ref{sec:averaging_ru}, and~\ref{sec:diagrammatics} are mostly reviews of the mapping from circuit to statistical mechanics models, but with an extra emphasis on the boundary conditions of the model. In Section~\ref{sec:Mapping}, we explain, starting from the statistical mechanics models, how to arrive at the path counting problem.







\subsection{Folding}
\label{sec:folding}

To perform the integral in Eq.~\eqref{eq:avg_fidelity}, it will be convenient to ``fold'' the circuit diagram. Recall that the expression for the fidelity is $F= \langle \psi_n | (|0\rangle\langle0|_1 \otimes I   ) |\psi_n\rangle$. This expression can be alternatively viewed as an inner product between two vectors:
\begin{equation}
\begin{aligned}
F &= \langle \psi_n |_{L} (|0\rangle_{1L}\langle0|_{1R} \otimes I   ) |\psi_n\rangle_{R}\\
&=  \langle \Psi| \left(|\psi_n\rangle^{\otimes 2}\right),
\end{aligned}
\label{eq:fidelity_fold}
\end{equation}
where 
\begin{equation}
    |\Psi\rangle = |0\rangle_{L} |0\rangle_{R} \otimes |\Phi\rangle_2 \otimes \ldots \otimes |\Phi\rangle_n
\end{equation}
is an unnormalized state with 
\begin{align}
    |\Phi\rangle_m = \sum_{i=1}^q |i\rangle_{mL} |i\rangle_{mR} \notag
\end{align}.



Eq.~\eqref{eq:fidelity_fold} can be alternatively understood via a folded circuit diagram. For instance, consider the circuit diagram shown in Fig.~\ref{fig:local_circuit}. This diagram represents the fidelity of the rewinding protocol applied to a local random quantum circuit. By folding this diagram, we can obtain the following expression for the fidelity:
\begin{align}
   \adjincludegraphics[width = 6.75 cm,valign=c]{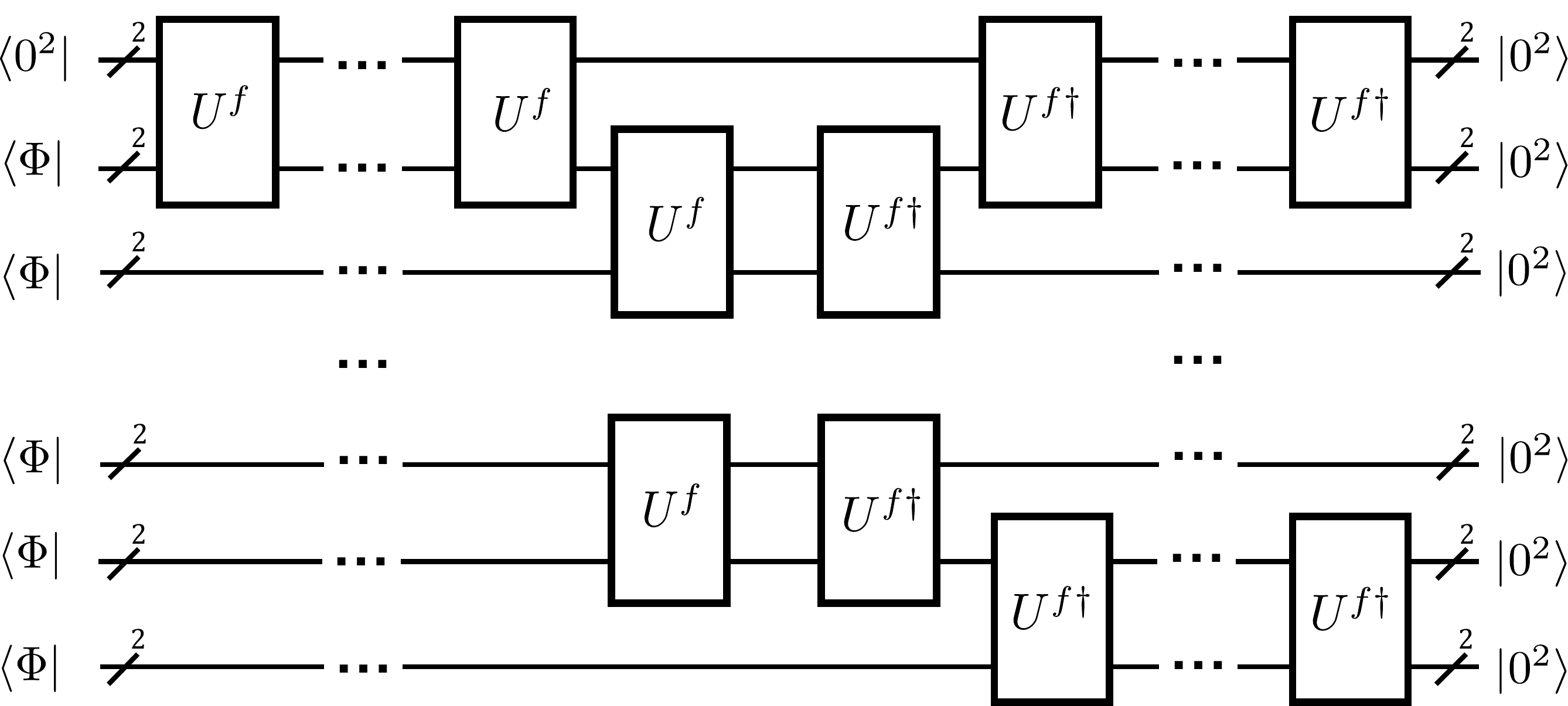}, \label{fig:single_fold}
\end{align}
where  $U^{f}_i$ is the {\it folded unitary}, defined as
\begin{align}
    U_{i}^{f} := U_{i}^{\dagger} \otimes U_{i}^{T}.
\end{align}
Notice that each leg of the folded unitary in fact consists of two legs, one belonging to $U_{i}^{\dagger}$ and the other to $U_{i}^{T}$.

We can now fold the circuit diagram one more time, obtaining the {\it doubly folded} diagram:
\begin{align}
   \adjincludegraphics[width = 6.75 cm,valign=c]{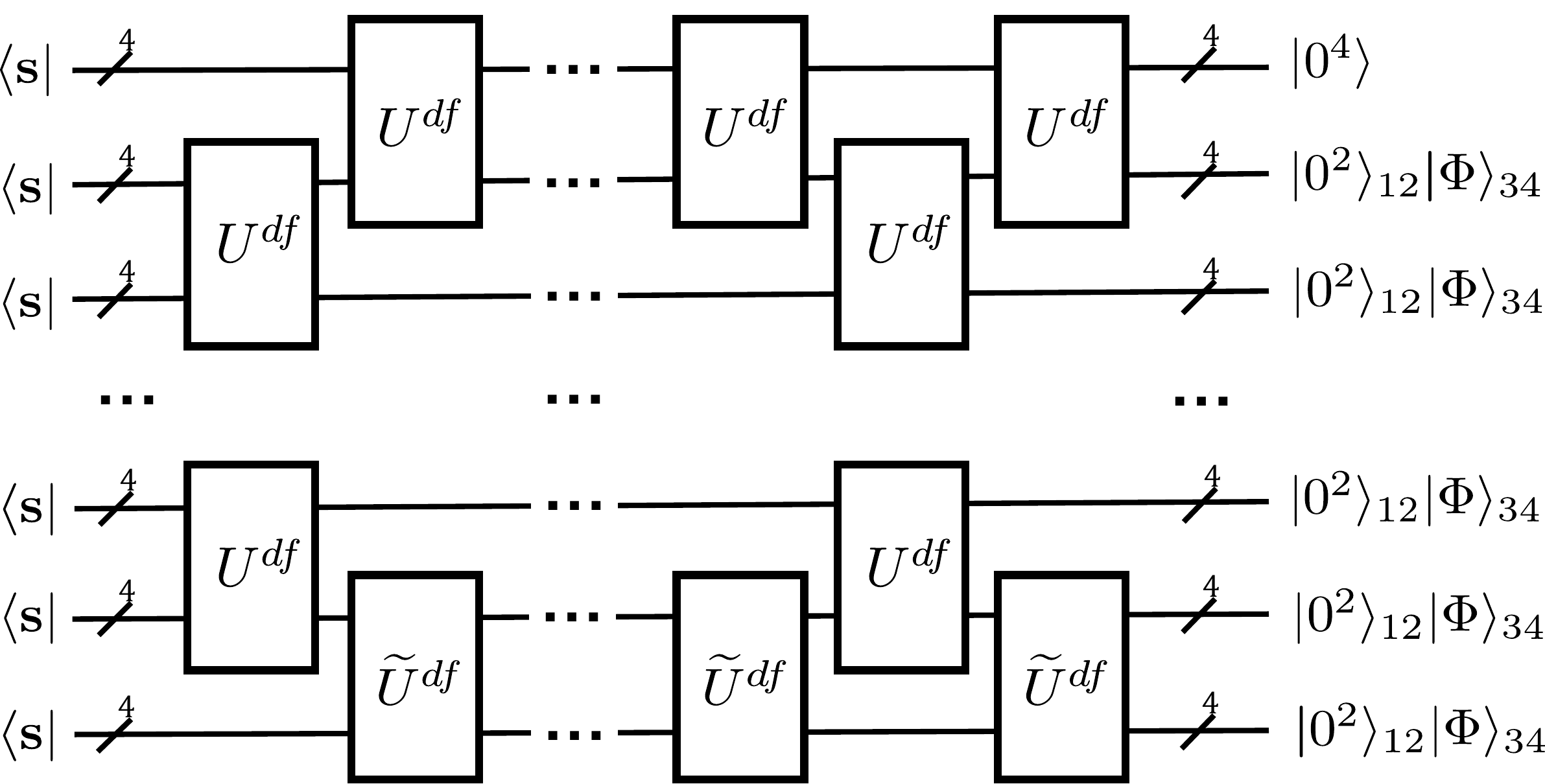}, \label{fig:doubly_fold}
\end{align}
where $| \mathbf{s} \rangle := |\Phi\rangle_{14} | \Phi \rangle_{23}$, $|\Phi\rangle_{ij} := \sum_{k=1}^{q} |k\rangle_i |k\rangle_j$. The unitaries shown in this diagram are the doubly folded unitaries, defined as
\begin{align}
    U^{df}_i &:= U_{i}^{*} \otimes U_{i} \otimes U_{i}^{*} \otimes U_{i} \notag \\ \widetilde{U}^{df}_i &:= U_{i}^{*} \otimes U_{i} \otimes I \otimes I. \label{eq:folding}
\end{align}

\subsection{Averaging Random Unitaries}
\label{sec:averaging_ru}

Our goal now is to evaluate the averaged fidelity, defined in Eq.~\eqref{eq:avg_fidelity}. Notice that, in the doubly folded diagram (see Eq.~\eqref{fig:doubly_fold}), the individual doubly folded unitary consists of four or two copies of identical unitaries. Moreover, by construction, the averaging of unitaries can be performed individually over each doubly folded unitary. 

Such calculation can be done conveniently using the so called Weingarten calculus~\cite{collins2003moments}, which provides moments of the unitaries over the Haar measure. The results related to this paper are listed below:
\begin{align}
    \int d \mu(U)\ U_{i_1 j_1} U_{i_1' j_1'}^* = \frac{1}{d} \delta_{i_1 i_1'} \delta_{j_1 j_1'} \label{eq:weingarten1}
\end{align}
and
\begin{align}
    &\quad \int d\mu(U)\ U_{i_1 j_1} U_{i_2 j_2} U_{i_1' j_1'}^* U_{i_2' j_2'}^* \notag \\ &= \sum_{\sigma, \tau \in S_2} \delta_{i_1 i_{\sigma(1)}'} \delta_{i_2 i_{\sigma(2)}'} \delta_{j_1 j_{\tau(1)}'} \delta_{j_2 i_{\tau(2)}'} Wg(\sigma \tau^{-1}, d) \label{eq:weingarten2}
\end{align}
where $S_2 = \{\mathbf{1}, \mathbf{s}\} \cong \mathbb{Z}_2$ is the premutation group over two elements. The expressions for the Weingarten functions are listed below.
\begin{align}
    Wg(1,d) = \frac{1}{d^2 - 1}, \quad Wg(s,d) = - \frac{1}{d(d^2-1)}
\end{align}

By applying these formulae, we obtain
\begin{align}
    &\int d\mu(U)\ \adjincludegraphics[width=2.2cm,valign=c]{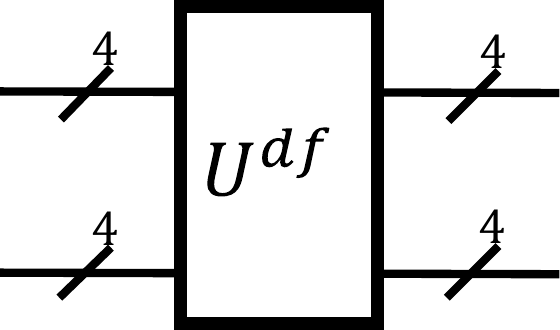}\notag \\ &= \sum_{\sigma,\tau \in S_{2}} Wg (\sigma \tau^{-1}, d)\ \adjincludegraphics[width=2.2cm,valign=c]{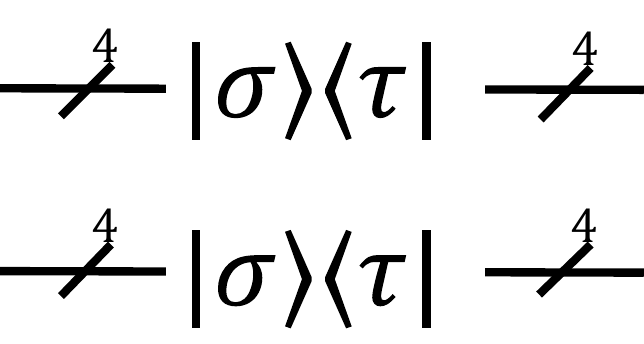}, \label{Wein_1}
\end{align}
where $|\sigma\rangle, |\tau\rangle \in \{|\mathbf{1}\rangle, |\mathbf{s}\rangle\}$ and $|\mathbf{1}\rangle = |\Phi\rangle_{12}|\Phi\rangle_{34}$ and $|\mathbf{s}\rangle = |\Phi\rangle_{14}|\Phi\rangle_{23}$, $|\Phi\rangle_{ij} = \sum_{k=1}^{q} | k \rangle_{i}| k \rangle_{j}$. Detailed calculations regarding the above relation can be found in Appendix~\ref{calculation}.

Also we have,
\begin{align}
    \int d\mu(U)\ \adjincludegraphics[width=2cm,valign=c]{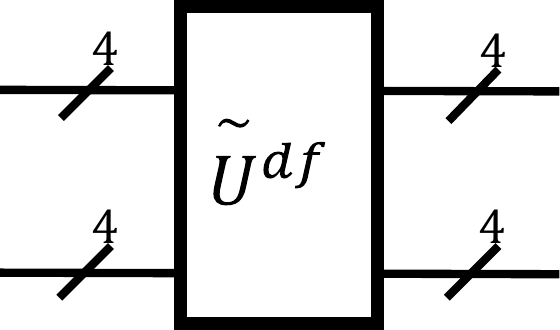} = \frac{1}{q^2}\ \adjincludegraphics[width=2.45cm,valign=c]{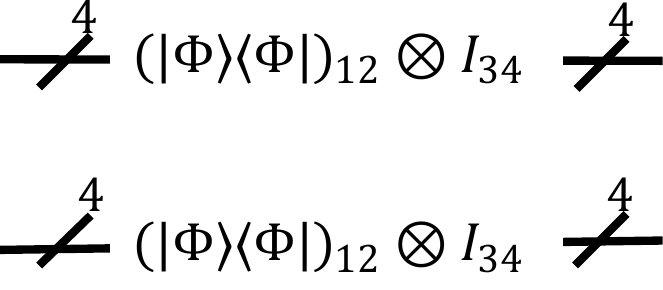}, \label{Wein_2}
\end{align}
and
\begin{align}
    \int d\mu(U)\ \adjincludegraphics[width=1.55cm,valign=c]{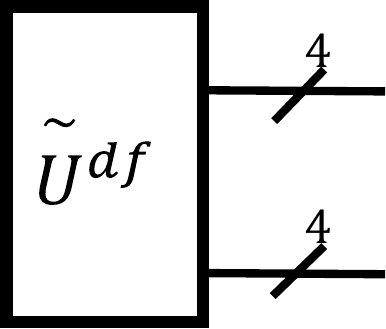} = \frac{1}{q^2}\ \adjincludegraphics[width=1.52cm,valign=c]{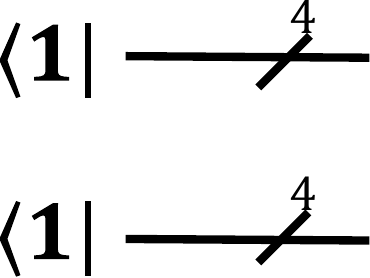}.
    \label{Wein_5}
\end{align}

\subsection{Diagrammatics}
\label{sec:diagrammatics}

In this section, we simplify the integrated random unitaries we get in Eq.~\eqref{Wein_2} and Eq.~\eqref{Wein_5} to a diagrammatic rule. This diagrammatic rule consists of trivalent diagrams (see Eq.~\eqref{wein_diag}), which we can multiply together to get the averaged fidelity.

We can partially contract the doubly folded unitaries in Eq.~\eqref{Wein_1} by $\langle \tau_1 |$ and $\langle \tau_2 | \in S_2$ from the left side. We define
\begin{align}
    Wg (\sigma \tau_3^{-1}, d)\ \adjincludegraphics[width=2.5cm,valign=c]{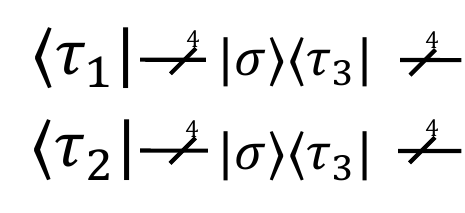} := \adjincludegraphics[width=2cm,valign=c]{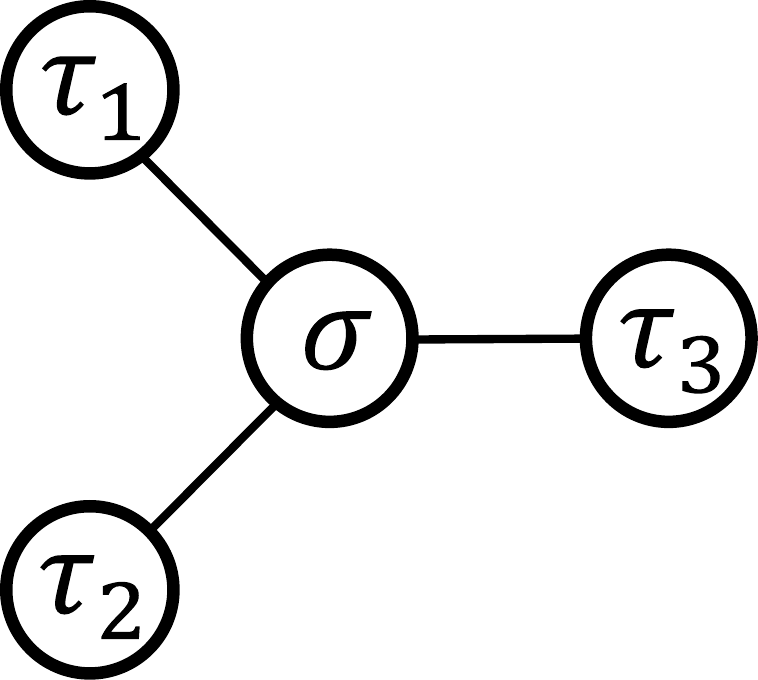}.
\end{align}
Then the contracted version of Eq.~\eqref{Wein_1} can be diagrammatically represented by~\cite{Nahum2018,Hunter-Jones2019}:
\begin{align}
     \int d\mu(U)\ \adjincludegraphics[width=2cm,valign=c]{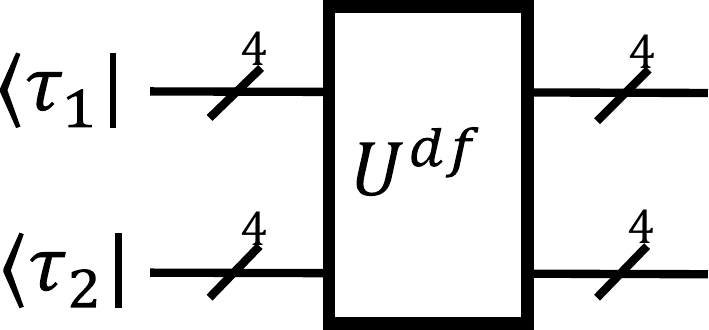} = \sum_{\sigma, \tau_3 \in S_2} \adjincludegraphics[width=2cm,valign=c]{diag_1.pdf}. \label{wein_diag}
\end{align}
By summing over the elements $\sigma \in S_2$, we can get rid of $\sigma$ in the middle and get
\begin{align}
    \adjincludegraphics[width=1.6cm,valign=c]{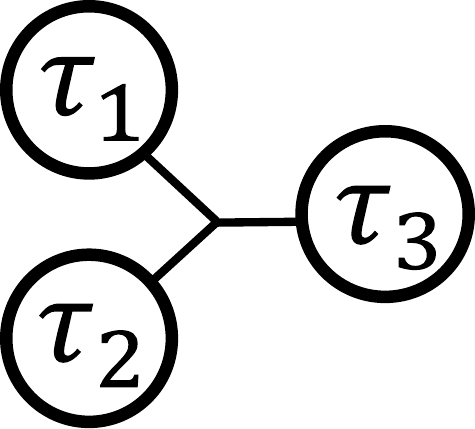} = 
    \begin{cases}
    1, \quad \tau_1 = \tau_2 = \tau_3 \\
    \frac{q}{1+q^2}, \quad \tau_1 \neq \tau_2 \\
    0, \quad \text{otherwise}
    \end{cases} \label{Wein_3}
\end{align}

Similar calculation can be applied to Eq.~(\ref{Wein_2}). We find no matter which state $\langle \tau | \in S_2$ we contract on the left, the output state is always $\langle \mathbf{1} |$. So diagrammatically, the integral in Eq.~\eqref{Wein_2} can be represented by
\begin{align}
    &\int d\mu(U)\ \adjincludegraphics[width=2.6cm,valign=c]{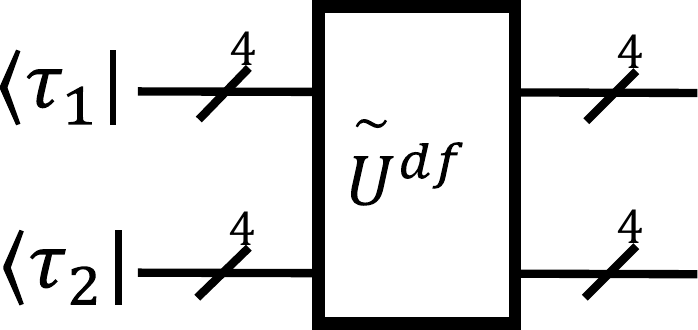} \notag \\&= \adjincludegraphics[width=1.6cm,valign=c]{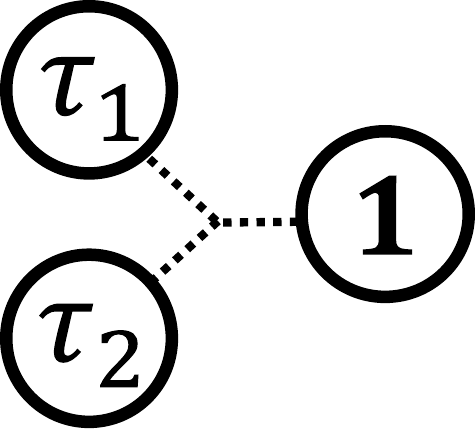} = 
    \begin{cases}
    1, \quad \tau_1 = \tau_2 = \mathbf{1} \\
    \frac{1}{q}, \quad \tau_1\ \text{or}\ \tau_2 = \mathbf{s}\\
    \frac{1}{q^2}, \quad \tau_1 = \tau_2 = \mathbf{s}
    \end{cases}. \label{Wein_4}
\end{align}
Here we use the dotted line to distinguish this diagram from the diagram in Eq.~\eqref{Wein_3}.

\subsection{Quantum circuits to path counting} \label{sec:Mapping}

Using the tools we reviewed in Section~\ref{sec:folding},~\ref{sec:averaging_ru}, and~\ref{sec:diagrammatics}, we can now relate the expression for the fidelity Eq.~\eqref{eq:avg_fidelity} to a path counting problem.

A path counting problem can be stated as following. Without loss of generality, consider a graph $G(V, E)$, where $V$ is the set of vertices and $E$ is the set of edges, respectively. The goal is to count the number of paths between two different vertices $v_1, v_2 \in V$. The particular path counting problem we consider comes with extra constraints, specifically, on the direction one can take along the path and the boundary conditions, on a square lattice.




To explain the key idea, it will be instructive to start with a concrete example. Consider, for instance, a simple doubly folded random quantum circuit as an example, constructed from  a local circuit with 6 initial qudits and 6 layers of Haar random gates. According to Eq.~\eqref{eq:avg_fidelity}, the average fidelity after we apply the rewinding protocol is given by the following integral:
\begin{align}
   \int  d\mu(U)\  \adjincludegraphics[width = 5 cm,valign=c]{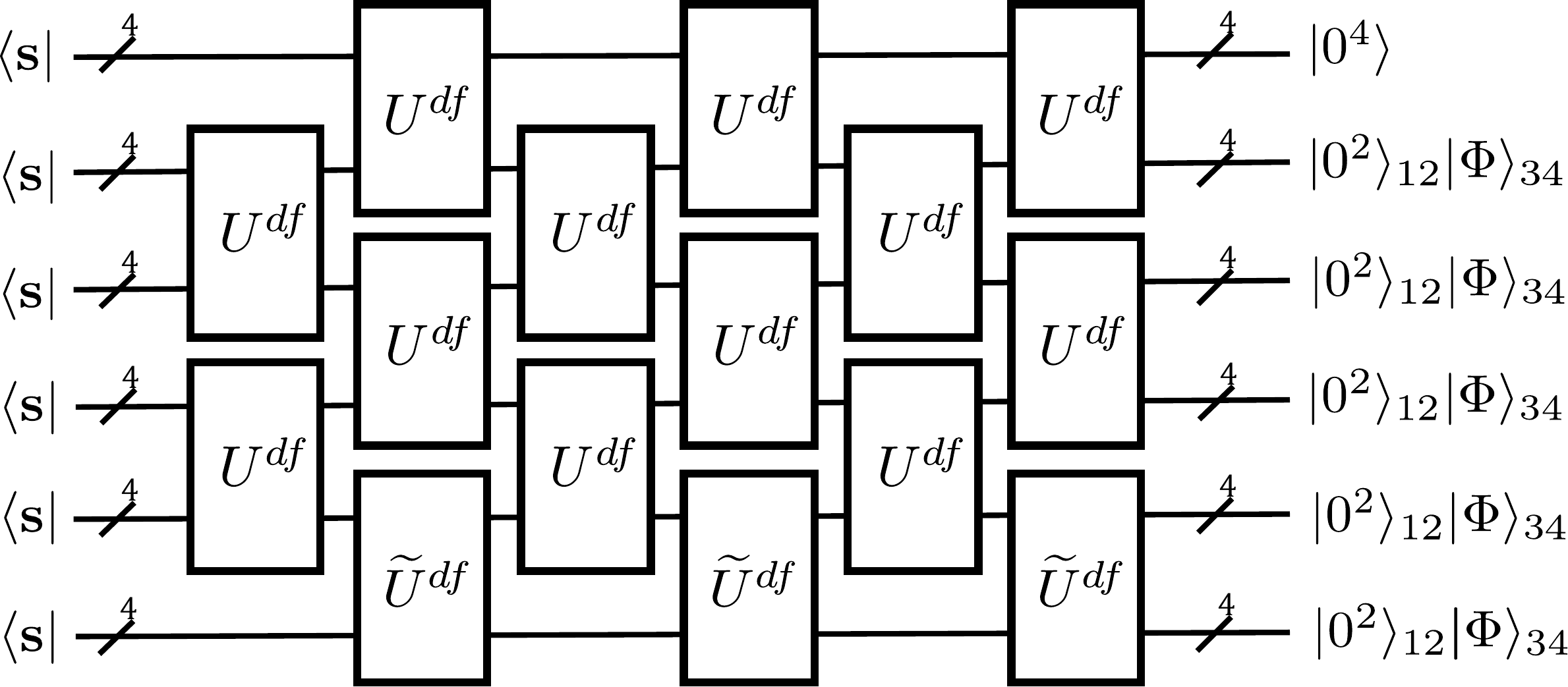}. \label{integral_example}
\end{align}
Diagrammatically, this integral can be represented by the following diagram according to the rules we introduce in Eq.~\eqref{Wein_3} and Eq.~\eqref{Wein_4}.
\begin{align}
    \adjincludegraphics[width = 6.75 cm,valign=c]{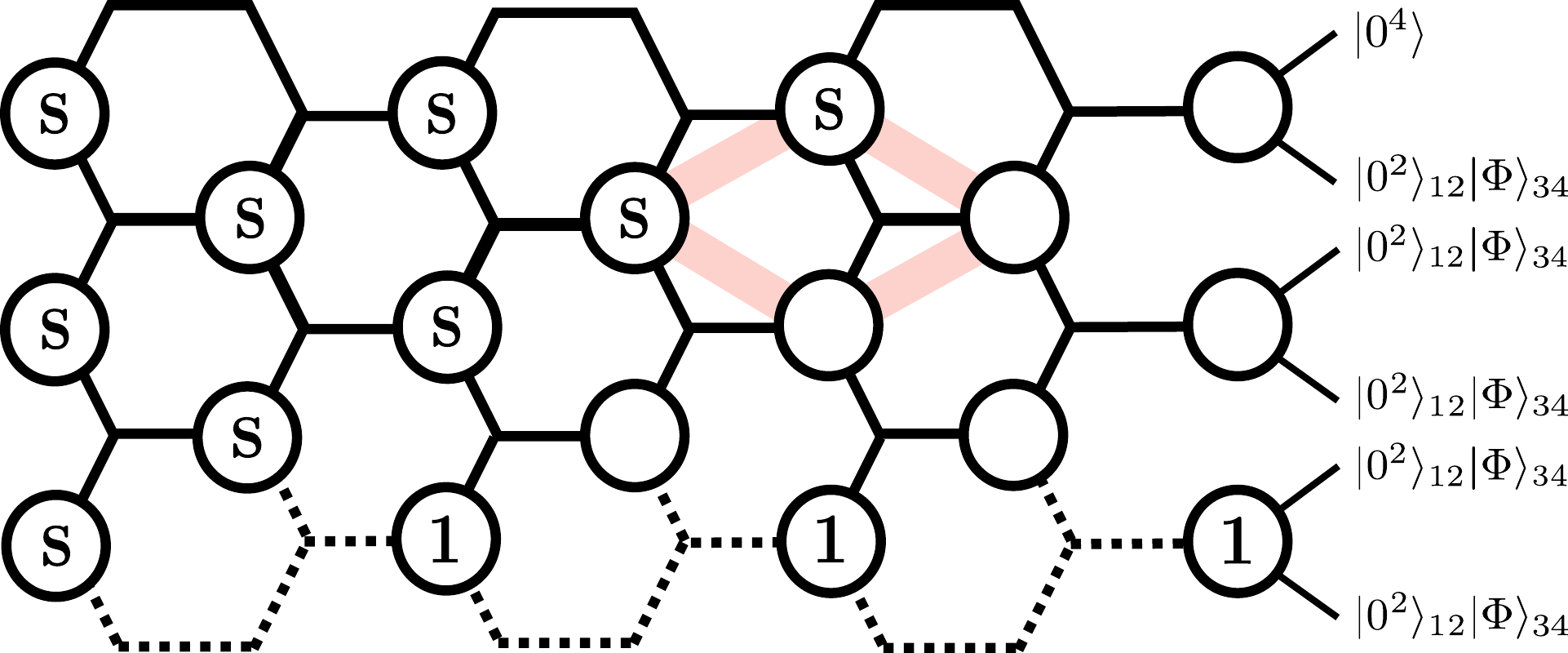}. \label{diagram_example}
\end{align}
We notice that the upper-left region and the boundary in the bottom are fixed according to the rules in Eq. (\ref{Wein_3}) and Eq.~(\ref{Wein_4}). Now we see the original random quantum circuit maps to a lattice model on a honeycomb sublattice. Each site can be either $\mathbf{s}$ or $\mathbf{1}$. The result of a single diagram in Eq.~\eqref{diagram_example} is given by the product of contributions from each trivalent term and the boundary conditions on the right side of the diagram. Eq.~\eqref{integral_example} is then equal to a sum of all possible partitions~\cite{Nahum2018,Hunter-Jones2019}.

However, further simplification is possible in our setup. 
Applying the rule derived in Eq.~\eqref{Wein_3}, we find that the top-left section yields the identity. Consequently, we begin by discarding this portion, resulting in the following diagram.
\begin{align}
\adjincludegraphics[width=6cm,valign=c]{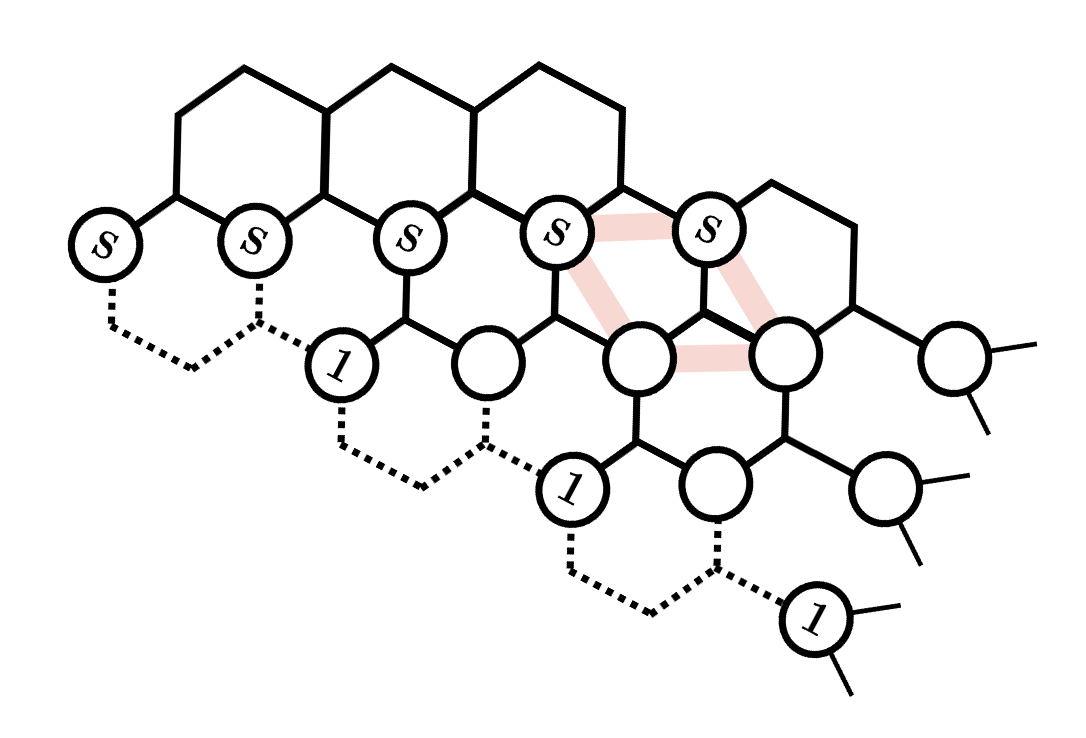}.
\end{align}
To simplify further, we reshape the rhombic region in this diagram into a square, yielding the diagram depicted below.
\begin{align}
    \adjincludegraphics[width = 6 cm,valign=c]{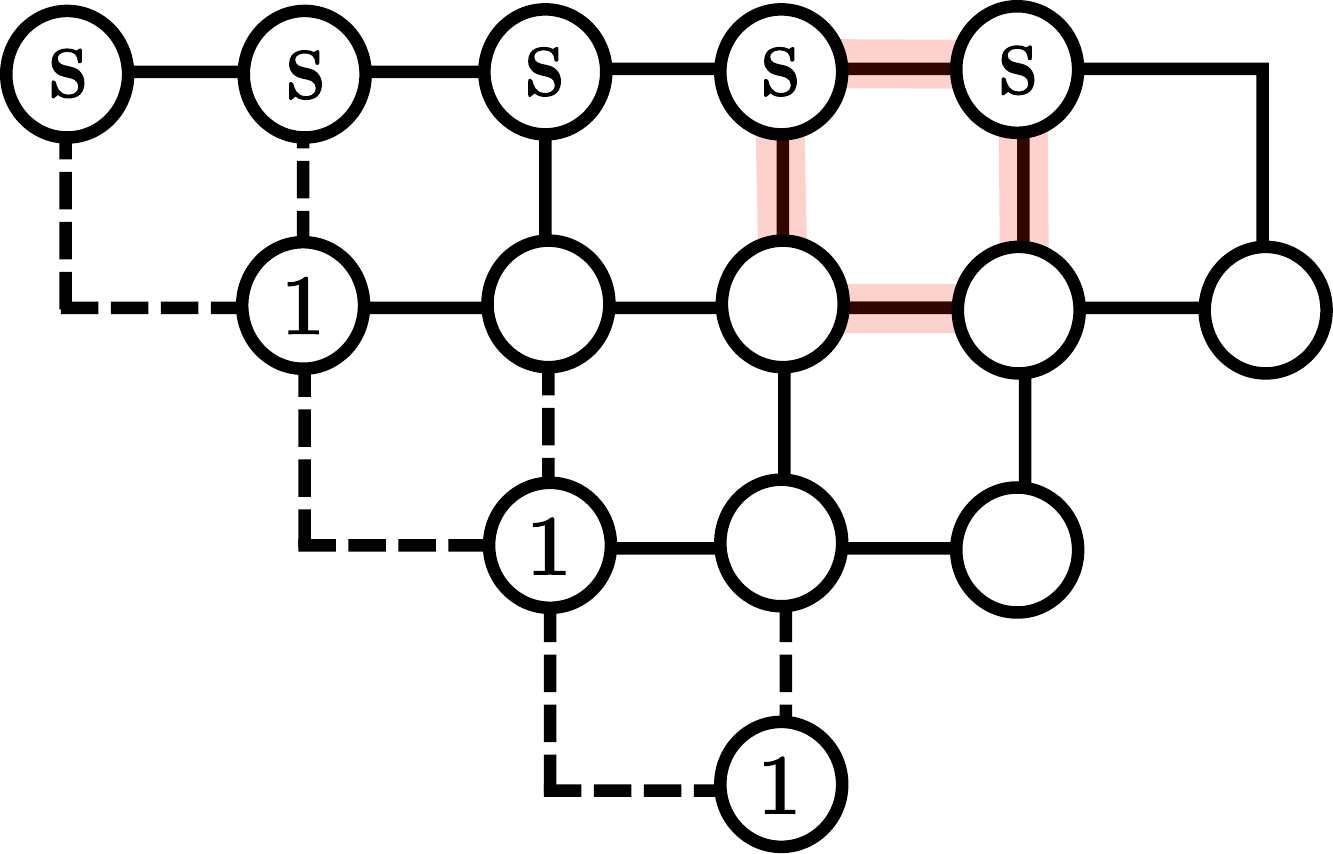} \label{diag_8}
\end{align}

We notice that the left boundary is composed of $\mathbf{1}$-nodes, while the upper boundary is composed of $\mathbf{s}$-nodes, except for the top-left corner. This distinctive boundary condition significantly limits the possible configurations of the graph. As a consequence, the system becomes divided by a domain wall, separating a pure $\mathbf{s}$ region from a pure $\mathbf{1}$ region.

\begin{figure*}
         \centering
         \includegraphics[width=14cm]{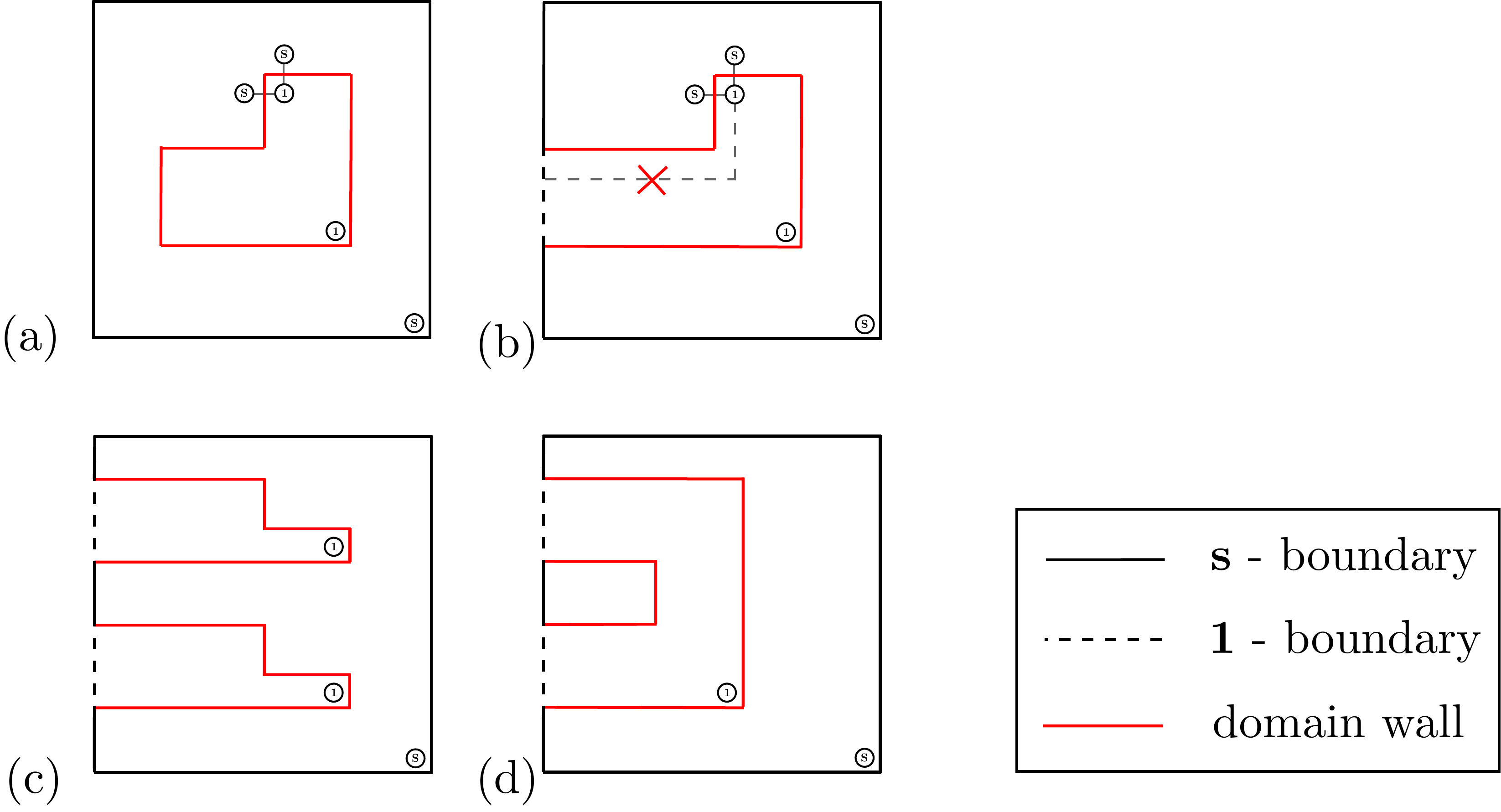}
         \caption{(a) The $\mathbf{1}$-region is isolated from the boundaries. One can always find a node that both the left and upward nodes are $\mathbf{s}$. (b) A similar situation happens when some of the $\mathbf{1}$-nodes are not connected to the $\mathbf{1}$-boundary by the directed paths. (c) When multiple disjoint $\mathbf{1}$-boundaries exist, the $\mathbf{1}$-regions can be disconnected. (d) When multiple disjoint $\mathbf{1}$-boundaries exist, the $\mathbf{1}$-regions can be simply connected. We notice that for both cases (c) and (d), there are two domain walls, which is equivalent to the number of disjoint $\mathbf{1}$-boundaries.}
         \label{fig:reply17}
\end{figure*}

We can argue this statement in two steps. First, we argue that the $\mathbf{1}$-regions must be connected to the left $\mathbf{1}$-boundaries, and for any $\mathbf{1}$-node, there must exist a directed path connecting it to a left $\mathbf{1}$-boundary, only moving upward and to the left, without intersecting any $\mathbf{s}$-nodes. If we consider a $\mathbf{1}$-region that is not attached to the left $\mathbf{1}$-boundaries or a region that prevents such paths, there must be a node within that region where both the left and upward nodes are $\mathbf{s}$. Based on Eq. (24), this configuration results in a total graph value of 0. Illustrative examples of these configurations are presented in Fig.~\ref{fig:reply17} (a) and (b).

In this case, as we have a single left $\mathbf{1}$-boundary, the only possible topology for the $\mathbf{1}$-regions that satisfies the aforementioned condition is a disc connected to the left $\mathbf{1}$-boundary. Consequently, there exists only one domain wall within the system. However, when multiple disjoint $\mathbf{1}$-boundaries are present on the left, we conjecture that the number of domain walls equals the number of disjoint $\mathbf{1}$-boundaries. The topology of the $\mathbf{1}$-region can exhibit greater complexity. Currently, our considerations suggest that the $\mathbf{1}$-region's topology can be either simply connected or not. Illustrative examples are presented in Fig.~\ref{fig:reply17} (c) and (d).

Therefore, the total number of different configurations of domain walls, which are also the boundary of $\mathbf{s}$ region, is given by the number of directed paths from the right undecided nodes on the boundary to the $\mathbf{s}$ on the upper-left corner. Moreover, if we start the path from the right boundary, the allowed directions it can go are only up and left.

As shown in Eq.~\eqref{diag_8}, the entire left boundary, except for the top left node in the corner, consists of $\mathbf{1}$-nodes. Consequently, the lattice model we are considering allows for only one domain wall between $\mathbf{1}$- and $\mathbf{s}$-regions, based on the aforementioned argument. The specific boundary condition of this model greatly simplifies the configurations of viable domain walls, making the analysis more tractable.

As for the weight of the domain wall, its weight contribution of a unit length of domain wall is $\frac{q}{1+q^2}$ in the bulk and $\frac{1}{q}$ when it is next to the $\mathbf{1}$-boundary. Diagrammatically, the unit weights can be represented by
\begin{align}
    \adjincludegraphics[width = 1.5 cm,valign=c]{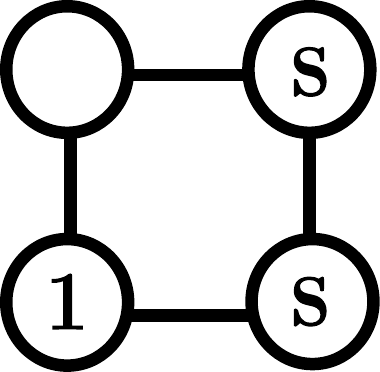} = \frac{q}{1+q^2},\quad \adjincludegraphics[width = 1.5 cm,valign=c]{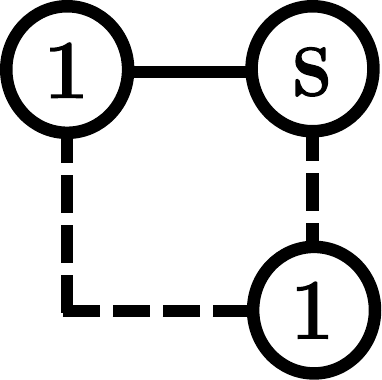} = \frac{1}{q}. \label{eq:unit_weights}
\end{align}
At the left end point of each domain wall, there is a overall factor of $\frac{1}{q^2}$, which in the diagram is represented by
\begin{align}
    \adjincludegraphics[width = 1.5 cm,valign=c]{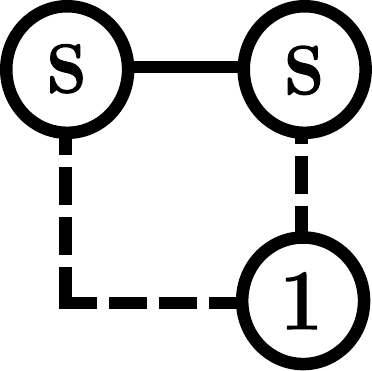} = \frac{1}{q^2}. \label{eq:end_weights}
\end{align}

Taking into account of all these contributions, we see that Eq.~(\ref{diag_8}) can be further simplified to the one shown in Eq.~\eqref{diag_10}. The green line (the bold, green line) is the boundary of the $\mathbf{s}$-region. The black points are the possible starting points of the paths. The red points (dark, gray points on the left) are the $\mathbf{1}$s that in Eq.~(\ref{diagram_example}) and Eq.~(\ref{diag_8}), which the paths are forbidden to go through. The blue point (light, gray point on the top-left corner) is the destination of every paths.
\begin{align}
    \adjincludegraphics[width = 5.5 cm,valign=c]{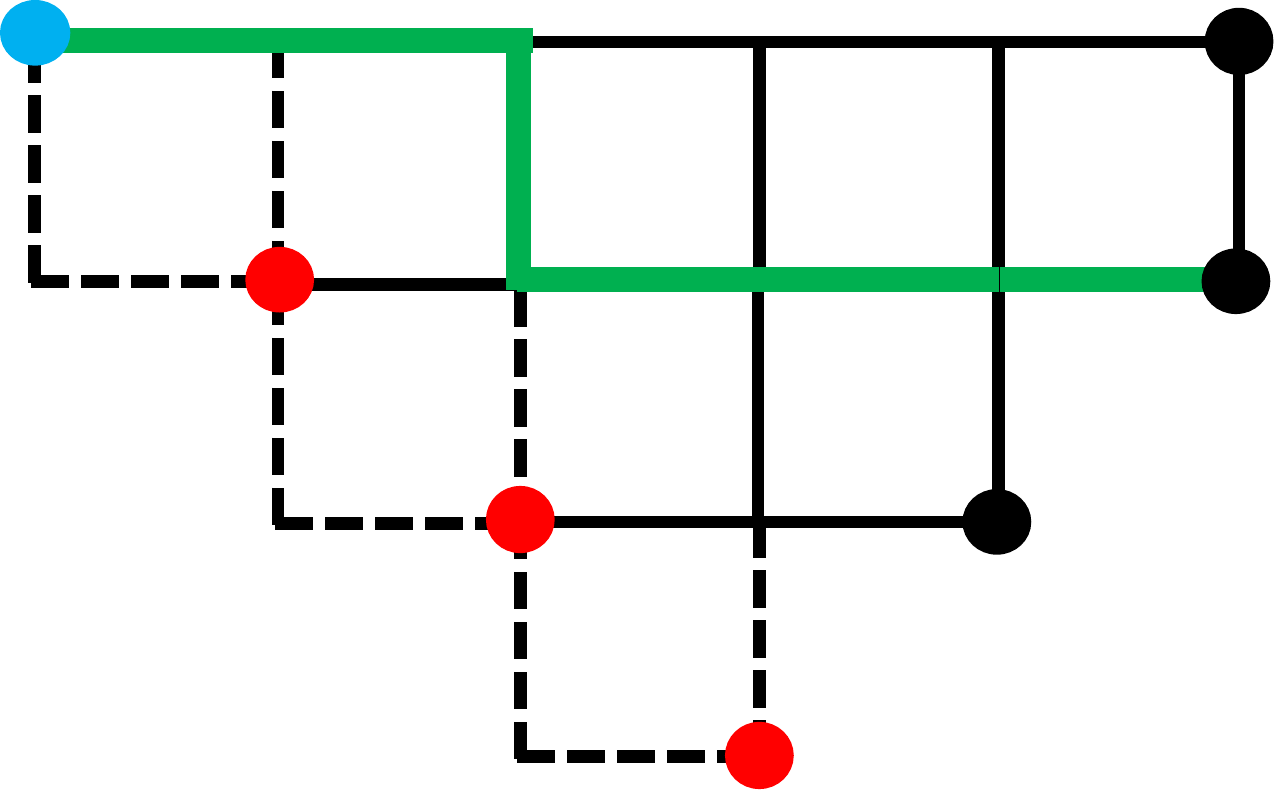} \label{diag_10}
\end{align}
For example, the weight of the green line in Eq.~(\ref{diag_10}) is 
\begin{align}
    \left(\frac{1}{q^2}\right) \left(\frac{1}{q}\right)  \left(\frac{q}{1+q^2}\right)^3 = \left(\frac{1}{1+q^2}\right)^3,
\end{align}
up to the boundary condition on the right as shown in Eq.~(\ref{diagram_example}).

The main lesson from this example is that not all domain walls need to be included in the calculation. In particular, it suffices to only consider configurations with a \emph{single} domain wall. Viewing this single domain wall as a path, we see that it suffices to consider the paths that satisfy the following conditions.
\begin{itemize}
    \item The path must start from one of the black dots on the right.
    \item Each step (from one vertex to the nearest-neighbor vertex) must  only go left or up.
    \item The path that goes through the red dots are forbidden.
    \item The path must end at the blue dot on the top-left corner.
\end{itemize}

The integral of Eq.~(\ref{integral_example}) is thus given by the sum of all weighted paths that satisfy the above conditions up to the boundary condition in Eq.~(\ref{diagram_example}).
\begin{align}
\boxed{
    F^{avg} = \sum_{\text{allowed paths}}\ (\text{weight})} \label{general_formula}
\end{align}
This is an important observation that is applicable to all the circuit family we consider. 

Remarkably, counting the number of these paths is equivalent to counting the number of directed paths on square lattices with linear boundaries is a problem already solved. We note the following theorems~\cite{krattenthaler2015lattice}.
\begin{widetext}
\begin{theorem}
Let $a+t \geq b \geq a+s$ and $c+t \geq d \geq c+s$. The number of all paths from $(a,b)$ to $(c,d)$ staying weakly below the line $y = x + t$ and above the line $y = x + s$ is given by
\begin{align}
    &|L((a,b)\to (c,d))|x+t \geq y \geq x+s | = \sum_{k \in \mathbb{Z}} \left(\begin{pmatrix}
    c + d - a - b \\
    c - a - k(t-s+2)
    \end{pmatrix} - \begin{pmatrix}
    c+d-a-b \\
    c - b - k(t-s+2) + t + 1
    \end{pmatrix}\right).
\end{align}
\end{theorem}
\end{widetext}

Here the term "weakly below" means that the path can go through the points on the boundaries. However, in practice this expression is not easy to calculate because it is an infinite sum and the combination may contain very large factorials. Fortunately, this expression can be simplified to a finite sum~\cite{krattenthaler2015lattice}.
\begin{widetext}
\begin{theorem}
Let $a+t \geq b \geq a+s$ and $c+t \geq d \geq c+s$. The number of all paths from $(a,b)$ to $(c,d)$ staying weakly below the line $y = x + t$ and above the line $y = x + s$ is given by
\begin{align}
    &|L((a,b)\to (c,d))|x+t \geq y \geq x+s | \notag \\ 
    &= \sum_{k=1}^{(t-s+1)/2} \frac{4}{t-s+2} \left(2 \cos \frac{\pi k }{t - s + 2}\right)^{c+d-a-b} \sin \left( \frac{\pi k (a - b + t + 1)}{t - s + 2}\right) \sin \left(\frac{\pi k (c - d + t + 1)}{t-s+2}\right). 
\end{align}
\label{theorem_2}
\end{theorem}
\end{widetext}
We use the simplified notation $L_{(a,b)}^{(c,d)}$ instead of $|L((a,b)\to (c,d))|x+t \geq y \geq x+s |$ in the remaining of the paper and we specify the upper and lower boundaries explicitly before the calculation to avoid ambiguity.

\section{Recycling with perfect gates}
\label{sec:recycling_perfect}
Now we are in a position to compute the fidelity of the rewinding protocol Eq.~\eqref{eq:avg_fidelity} using the correspondence Eq.~\eqref{general_formula}. We consider the convolutional circuit, a hybrid circuit, and the local circuit.

\subsection{Convolutional circuits} \label{sec:convolutional_circuits}

A convolutional circuit is the one which can be written in the following form:
\begin{align}
    U_{[n-k+1],n} \cdot \cdot \cdot U_{[2,k+1]} U_{[1,k]}.
\end{align}
In this paper, we consider the convolutional circuits with 2-site Haar random gates shown in Fig.~\ref{fig:interpolation} (a).

The doubly folded diagram of the convolutional circuit is given by
\begin{align}
    \adjincludegraphics[width=6.5cm,valign=c]{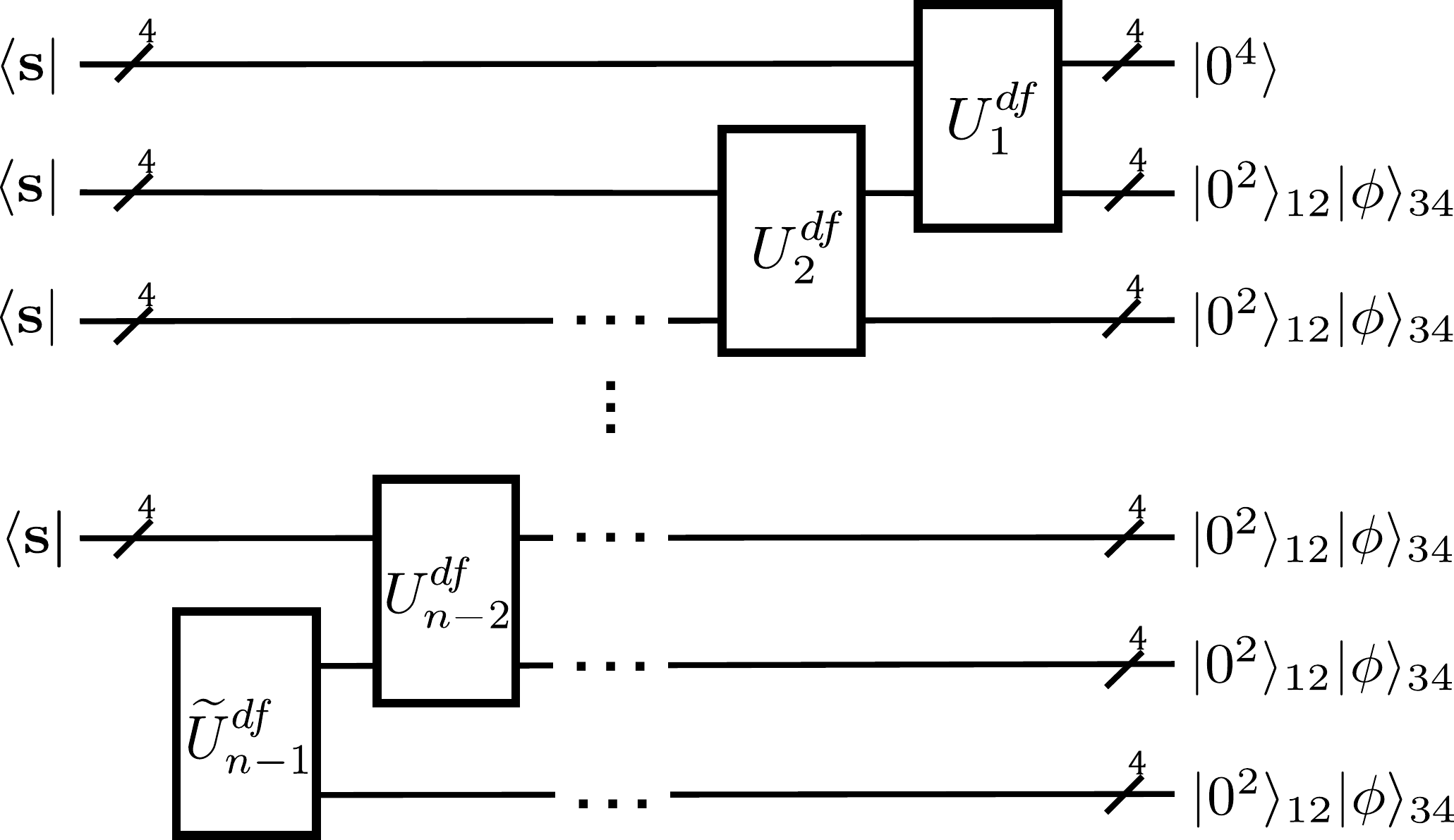}
\end{align}
The node diagram is given by
\begin{align}
    \adjincludegraphics[width=6.5cm,valign=c]{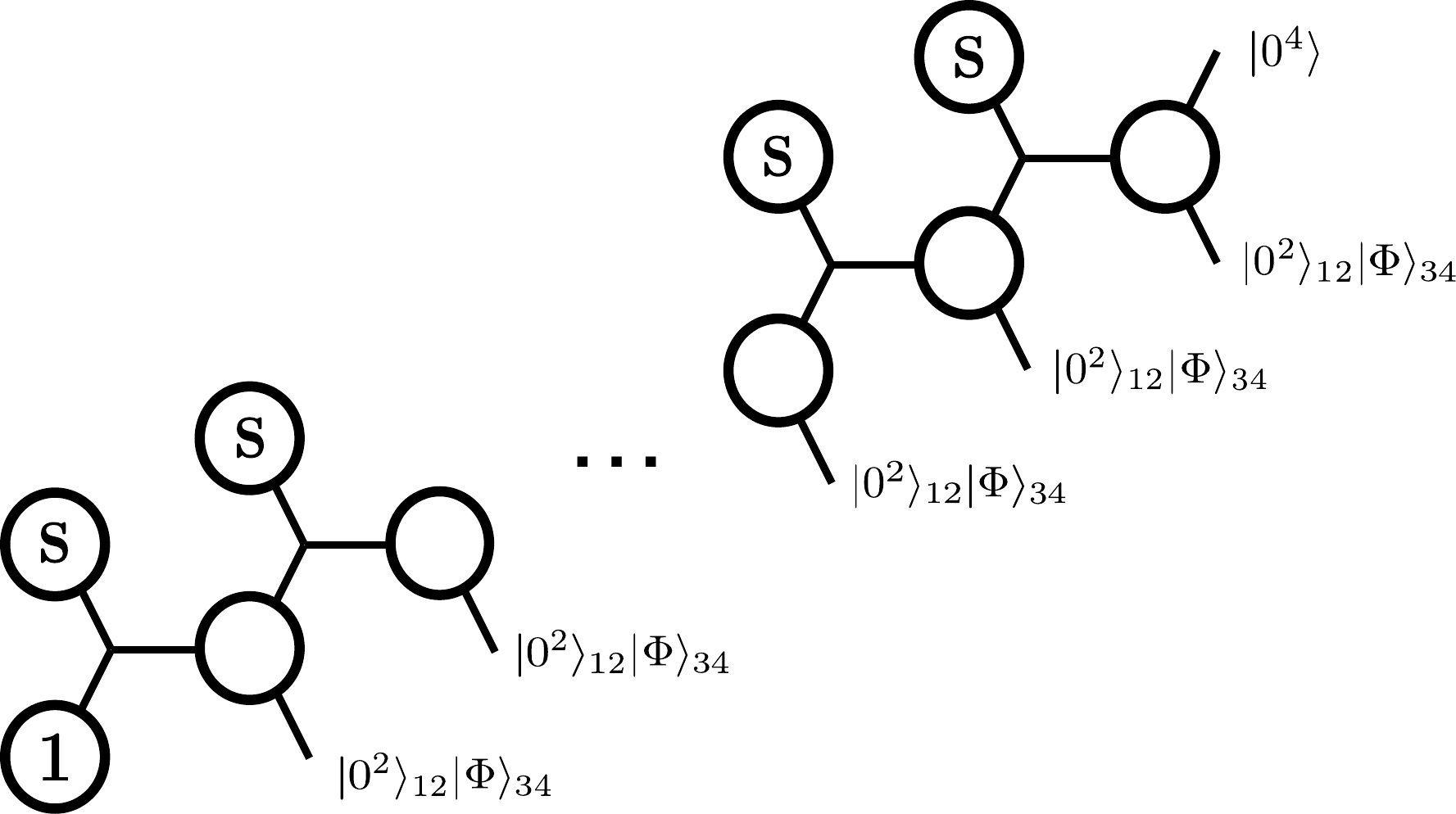} \label{lattice_node}
\end{align}
Thus, the average fidelity can be computed by counting paths in the following diagram.
\begin{align}
    \adjincludegraphics[width=6.5cm,valign=c]{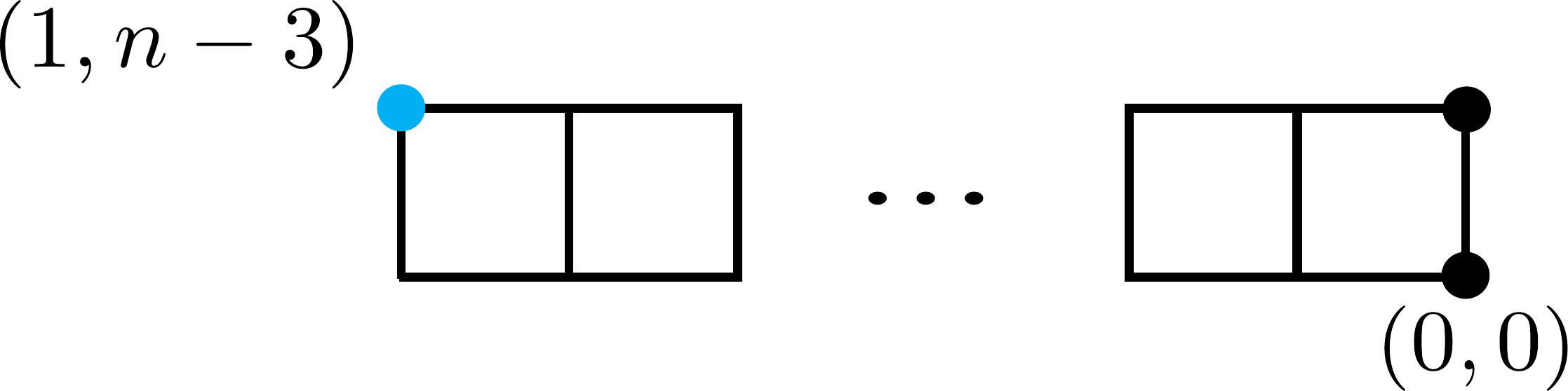}
\end{align}
By weighting each path appropriately and summing them over, the average fidelity becomes
\begin{align}
    F_1 &= \frac{1}{1+q^2} + \frac{1}{1+q^2} \left(\frac{q^2}{1+q^2}\right) + \frac{1}{1+q^2} \left(\frac{q^2}{1+q^2}\right)^2 \notag \\
    &+ ... + \frac{1}{1+q^2} \left(\frac{q^2}{1+q^2}\right)^{n-3} + \frac{1}{q} \left(\frac{q^2}{1+q^2}\right)^{n-2} \notag \\
    &=1 - \frac{q-1}{q} \left(\frac{q^2}{q^2 + 1}\right)^{n-2} \label{F1}
\end{align}
The diagram corresponding to recycling the first $k$ qudits has different boundary conditions with Eq.~(\ref{lattice_node}). The first $k$ states on the right boundary become $|0^4\rangle$. Therefore, the fidelity is given by
\begin{align}
    F_{1\to k} &= 1 - \left(\frac{q^2}{q^2+1}\right)^{n-k} \left(1 - \vphantom{\left(\frac{q}{q^2+1}\right)^{k-2}}
    \frac{1}{q(q^2-q+1)} \right. \notag \\ &\quad \left.\times \left(1 + \frac{(q-1)^2}{q}\left(\frac{q}{q^2+1}\right)^{k-2}\right)\right)
\end{align}
We notice that to get the same fidelity as Eq.~(\ref{F1}), one is expected to have a larger value of $n$ as $k$ is increasing.

For recycling the $i$-th ($n>i>1$) qudit, the corresponding diagram is obtained by swapping the first and $i$-th states on the right boundary. The fidelity is then given by
\begin{align}
    F_{i} = 1 - \frac{q-1}{q} \left(\frac{q^2}{q^2+1}\right)^{n-i}.
\end{align}
Similarly, by changing the corresponding boundary states, the fidelity of recycling the $i$-th and $j$-th ($i>j$) qudits is given by
\begin{align}
    F_{ij} = 1 - \frac{q-1}{q} \left(\frac{q^2}{q^2+1}\right)^{n-i}\left(1 + \frac{1}{q}\left(\frac{q^2}{q^2+1}\right)^{i-j}\right).
\end{align}
Therefore, the correlation function of fidelity between the $i$-th and $j$-th qudits is given by
\begin{align}
    &F_{ij}^c =F_{ij} - F_{i}F_{j} \notag \\&=  \left(\frac{q-1}{q}\right)^2\left(\frac{q^2}{q^2+1}\right)^{n-j} \left(1-\left(\frac{q^2}{q^2+1}\right)^{n-i}\right), \label{eq:noiseless_correlation}
\end{align}
which vanihes exponentially in $n$, as expected.

\subsection{Hybrid Circuits} \label{sec:hybrid_circuits}

\begin{figure*}
    \centering
    \includegraphics[width=16cm]{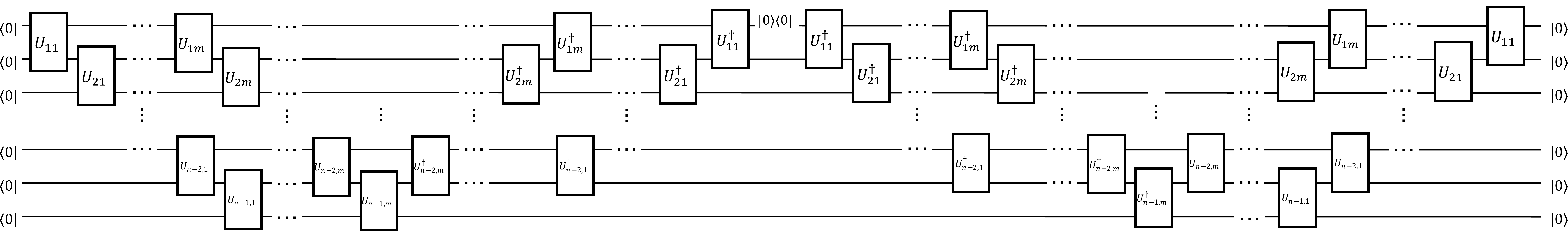}
    \caption{A diagrammatic representation of the fidelity of recycling the first qudit after applying an $m$-layers convolutional circuit with rewinding protocol. }
    \label{hybrid_circuit}
\end{figure*}

Diagrammatically, the hybrid circuits with $m$-layers of convolutional circuits are shown in Fig.~\ref{fig:interpolation} (c). The fidelity of recycling the first qudit is given by the circuit in Fig.~\ref{hybrid_circuit}.

The corresponding doubly folded diagram is the following:
\begin{align}
    \adjincludegraphics[width=6.5cm,valign=c]{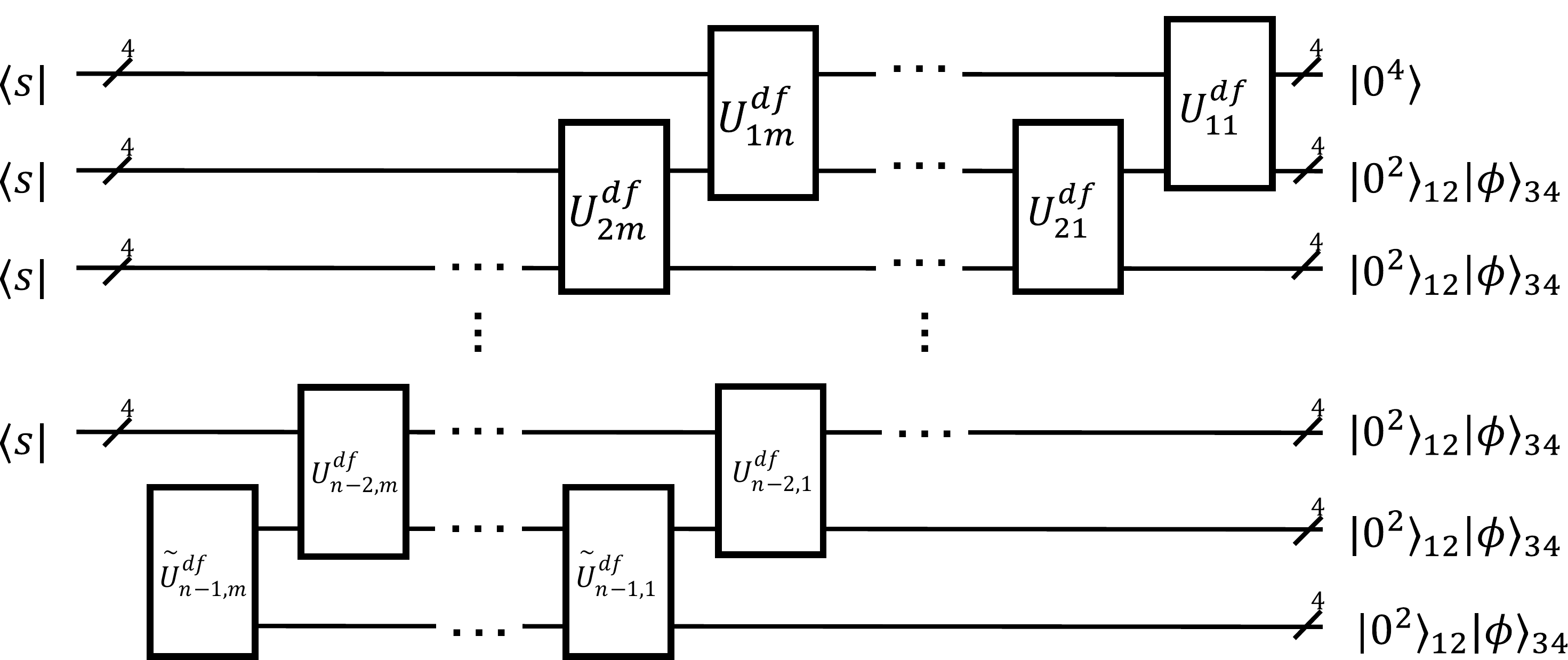} \label{fig:hybrid_df}
\end{align}

The lattice diagram is given by
\begin{align}
    \adjincludegraphics[width=7cm,valign=c]{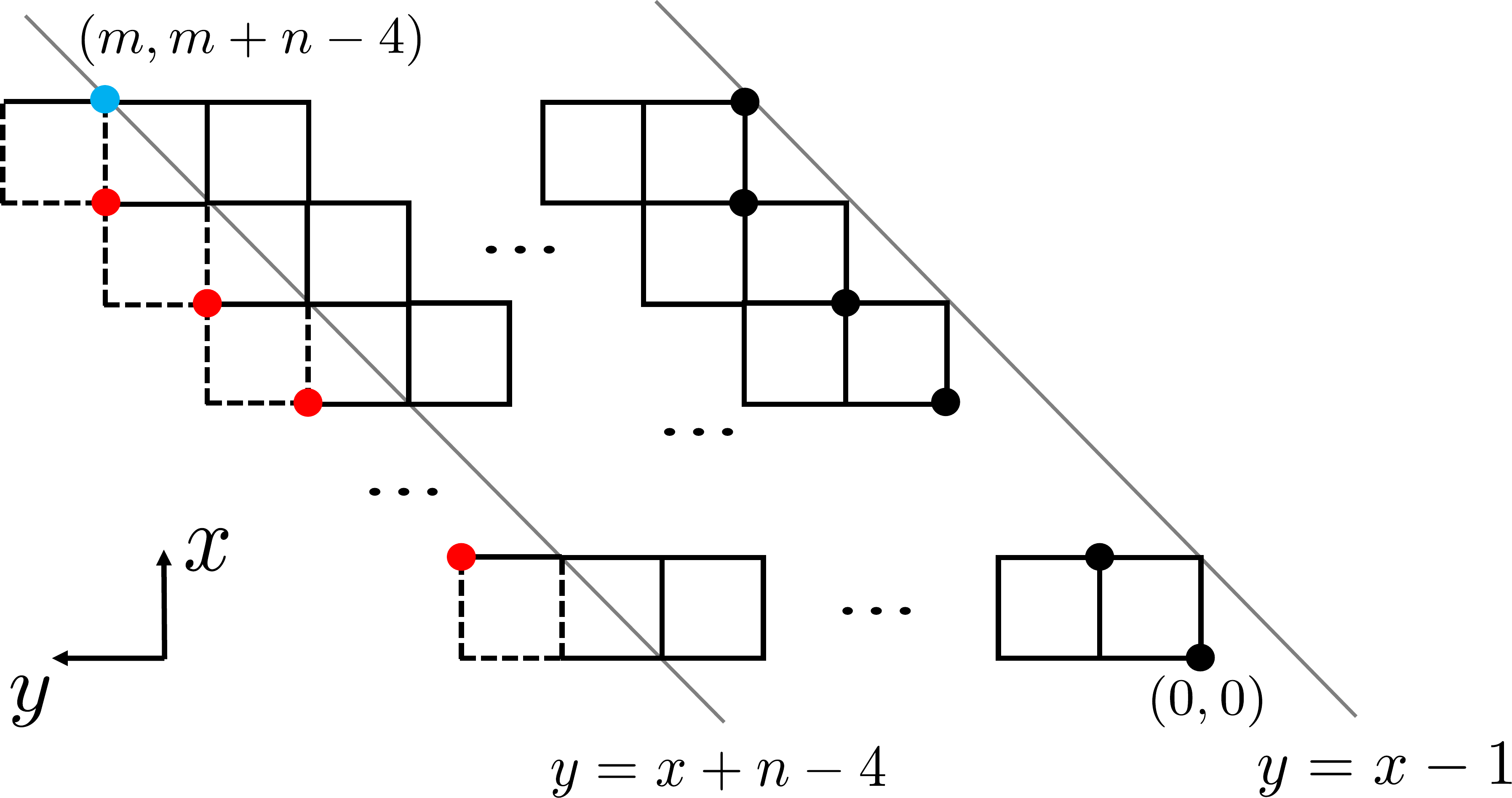} \notag
\end{align}
We first consider the case that all the nodes are $\mathbf{1}$s. The contribution from this case is given by
\begin{align}
    \frac{1}{q} \left(\frac{q^2}{1+q^2}\right)^{n-2}
\end{align}
Then we consider the case that some nodes on the right boundary $y = x$ to be $\mathbf{s}$, except for the point $(0, 0)$. 

The sum of weighted paths that do not touch the $y = x + n -4$ line and the $y = x - 1$ line is given by
\begin{align}
    \sum_{k = 1}^{n - 3} \frac{1}{q^{2m-2k+1}}\left(\frac{q^2}{1 + q^2}\right)^{n + 2m - 2k - 2} L_{1\ (k,k)}^{m,m+n-4}
\end{align}
Here we define $L_{1\ (k,k)}^{(m,m+n-4)}$ to be $| L ((k,k) \to (m, m+n-4)) | x + n - 5 \geq y \geq x - 1 |$ according to the definition in Theorem~\ref{theorem_2}.

We also need to consider the case that the paths go through the $y = x + n - 4$ line. Consider there are $l$ points of a path that are on the line. We define the $x$-value of these points as $\mathcal{I}_l = \{i_1, i_2, ..., i_l\}$ and the destination as $i_{l+1}$. A path that has $l$ nodes on the $y = x + n - 4$ line gets a factor $((1 + q^2)/q^2)^l$ on its weight. So the contribution from the paths that touch the left line is given by
\begin{align}
    &\frac{1}{q^{2m-2k+1}}\left(\frac{q^2}{1 + q^2}\right)^{n + 2m - 2k - 2}\sum_{l = 1}^{m - k} \left(\frac{1+q^2}{q^2}\right)^l \notag \\
    &\times \sum_{\mathcal{I}_l}\left( L_{2\ (k,k)}^{(i_1,i_1+n-4)}\prod_{j=1}^{l} L_{2\ (i_{j}+1,i_{j}+n-3)}^{(i_{j+1},i_{j+1}+n-4)}\right),
\end{align}
in which we define $L_{2\ (k,k)}^{(i_1,i_1+n-4)}$ to be $| L ((k,k) \to (i_1,i_1+n-4)) | x + n - 4 \geq y \geq x - 1 |$ and $\sum_{\mathcal{I}_l}$ to be the sum over all possible partitions of these $l$ points on the $y = x + n - 4$ line.

Similarly, we consider the node $(0, 0)$ to be $\mathbf{s}$. The paths can also be divided into two parts. One that touches the left line and the other does not. However, the calculation of this part is more complicated since we also need to consider the boundary condition on the bottom line, which are the $|0^2\rangle_{12} |\Phi\rangle_{34}$ states in Eq.~\eqref{fig:hybrid_df}.
The general formula for recycling the first qudit in the $m$-layers convolutional circuit is given by
\begin{widetext}
{\footnotesize
\begin{align}
    F_{1}^m &= \sum_{k=1}^{m-1} \left(\frac{q}{1+q^2}\right)^{n+2m-2-2k} q^{n-3}\left(L_{1\ (k,k)}^{(m,m+n-4)}+\sum_{l=1}^{m-k}\left(\frac{1+q^2}{q^2}\right)^l \sum_{\mathcal{I}_l} \left( L_{2\ (k,k)}^{(i_1,i_1+n-4)}\prod_{j=1}^{l} L_{2\ (i_{j}+1,i_{j}+n-3)}^{(i_{j+1},i_{j+1}+n-4)}\right)\right) \notag \\
    &+ \sum_{k=0}^{n-3} \frac{1}{q^{2m}} \left(\frac{q^2}{1+q^2}\right)^{n+2m-4-k}\left(L_{1\ (1,k)}^{(m,m+n-4)}+\sum_{l=1}^{m-1}\left(\frac{1+q^2}{q^2}\right)^l \sum_{\mathcal{I}_l}\left( L_{2\ (1,k)}^{(i_1,i_1+n-4)}\prod_{j=1}^{l} L_{2\ (i_{j}+1,i_{j}+n-3)}^{(i_{j+1},i_{j+1}+n-4)}\right)\right) \notag \\
    &+ \frac{1}{q} \left(\frac{q^2}{1+q^2}\right)^{n-2}
\end{align}}
\end{widetext}

While this general expression is undoubtedly complicated, it is interesting to plug in different values of $m$. For $m=1, n \geq 4$, we get
\begin{align}
    F_{1}^{m=1} = 1 - \frac{q-1}{q} \left(\frac{q^2}{q^2 + 1}\right)^{n-2},
\end{align}
which is exactly Eq. (\ref{F1}). For $m=2,n \geq 5$, we get
\begin{align}
    F_{1}^{m=2} = 1 - \frac{q - 1}{q} \left(\frac{q^2}{q^2 + 1}\right)^{n} \left(1+\frac{n}{q^2}+\frac{2}{q^4}\right)
\end{align}
For $m=3, n \geq 6$, We get
\begin{align}
    F_{1}^{m=3} &= 1 - \frac{q - 1}{q} \left(\frac{q^2}{q^2 + 1}\right)^{n+2}\left(1+\frac{1}{q^2}\right.\notag \\ &+\left.\frac{(1+n)(2+n)}{2q^4}+\frac{2(2+n)}{q^6}+\frac{3}{q^8}\right)
\end{align}
Importantly, for these finite values of $m$, we find the infidelity of these circuits to decay exponentially as the number of qudits increases. Interestingly, the decay rate remains a constant even with different numbers of layers $m$.

\begin{figure*}
     \centering
     \begin{subfigure}[b]{0.49\textwidth}
         \centering
         \includegraphics[width=\textwidth]{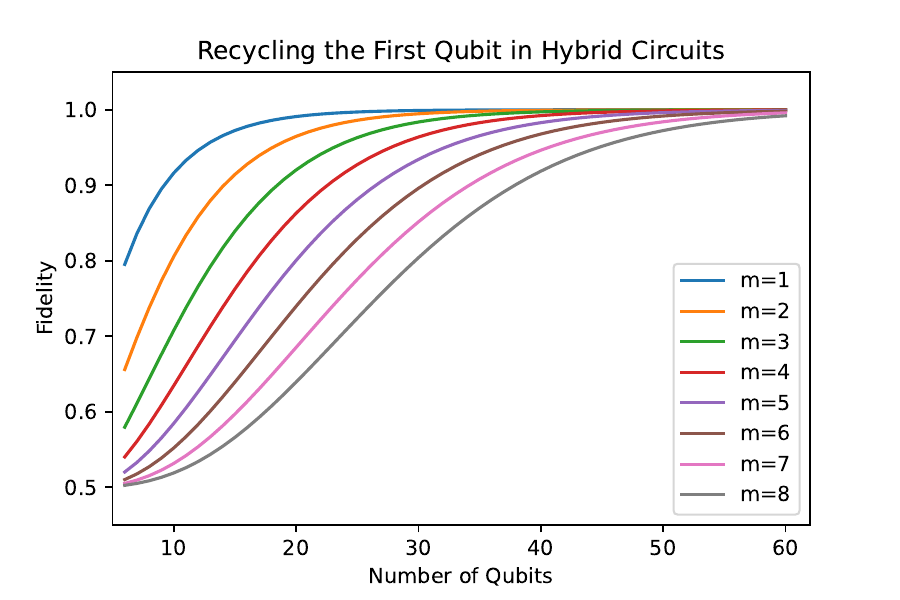}
         \caption{}
         \label{num_hybrid}
     \end{subfigure}
\hfill
     \begin{subfigure}[b]{0.49\textwidth}
         \includegraphics[width=\textwidth]{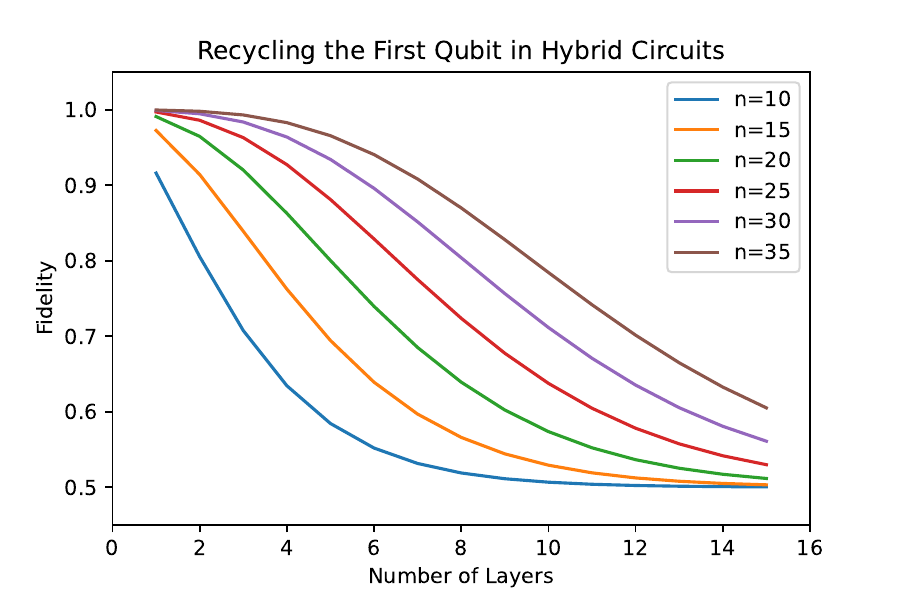}
         \caption{}
         \label{num_hybrid_1}
     \end{subfigure}\\
     \begin{subfigure}[b]{0.49\textwidth}
         \centering
         \includegraphics[width=\textwidth]{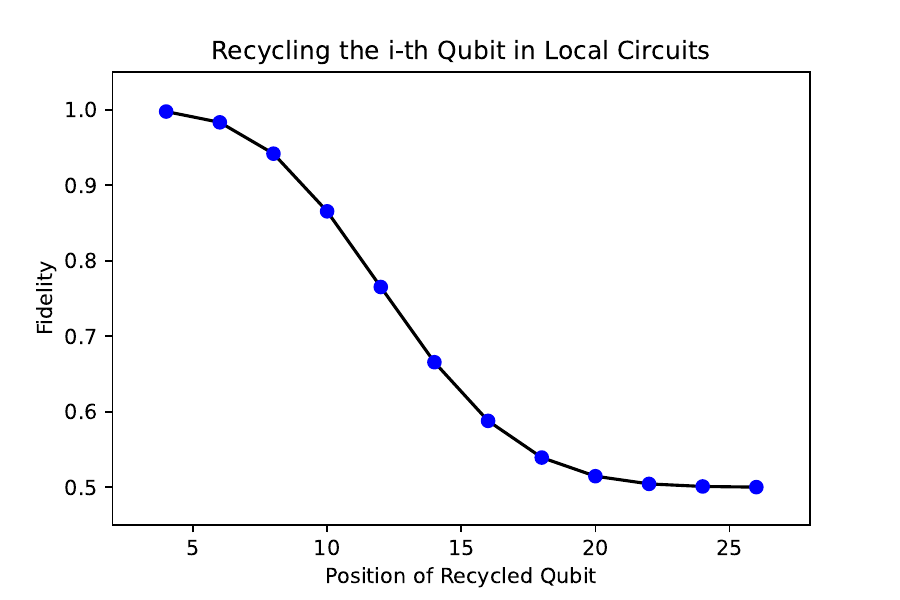}
         \caption{Local Circuits ($n > m$)}
         \label{fig:three sin x}
     \end{subfigure} 
     \hfill
     \begin{subfigure}[b]{0.49\textwidth}
         \includegraphics[width=\textwidth]{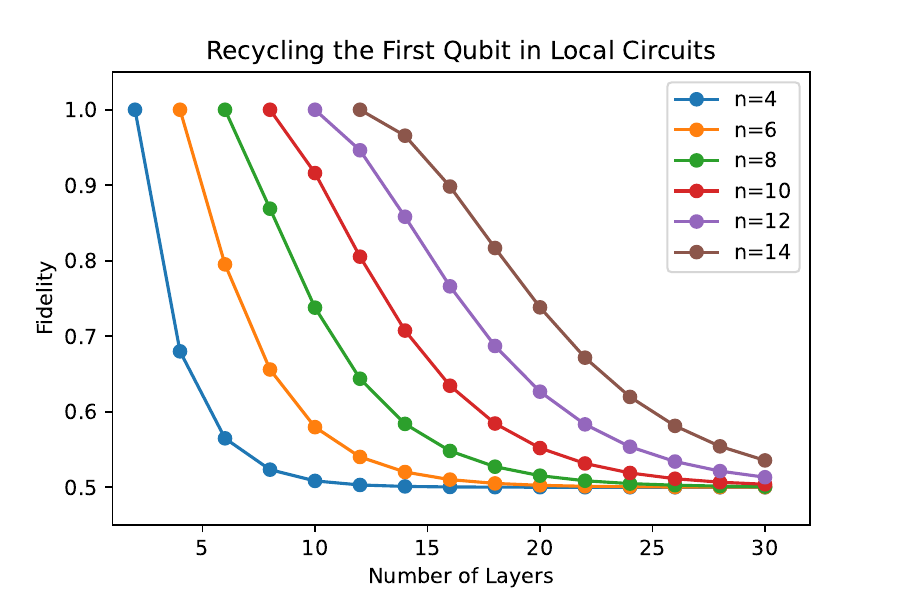}
         \caption{Local Circuits ($n < m$)}
         \label{num_local}
     \end{subfigure}
        \caption{The fidelity of recycling the first qubits in hybrid and local circuits. (a) For a given number of layers $m$, the infidelity exhibits exponential decay as $n$ increases. The top curve corresponds to the fidelity when $m=1$, and the curves below it represent increasing values of $m$ from top to bottom. (b) For a specific number of qubits $n$, the fidelity also decays exponentially as $m$ increases. The bottom curve represents the fidelity when $n=10$, and the curves above it correspond to increasing values of $n$ from bottom to top. (c) The fidelity of recycling the $i$-th qubit in a local circuit is depicted when $n > m$. Notably, as the position of the recycled qubit approaches the qubit in use, the fidelity drops to $1/2$ ($1/q$ for $q$-dimensional qudits). (d) The fidelity of recycling the first qubit in local circuits is shown when $m > n$. It is observed that as the number of layers $m$ increases, the fidelity drops to $1/2$ ($1/q$ for $q$-dimensional qudits). The leftmost curve illustrates the fidelity when $n=4$, with increasing values of $n$ represented from left to right.}
        \label{fig:hybrid_numerical}
\end{figure*}

One might also ask, once the total number of qudits $n$ is fixed, how the fidelity changes as $m$ increases. While we do not have a succinct expression for general $n$, a numerical calculation indicates that the fidelity decays exponentially with $m$; see Fig.~\ref{fig:hybrid_numerical}. This result is consistent with the analytical calculation for a specific $n$. When $n = 3$, which is the least number of qudits that can make the convolutional circuits possible, the fidelity is given by
\begin{align}
    F_{1}^{n=3} = \frac{1}{q} + \frac{q-1}{q} \left(\frac{1}{1+q^2}\right)^m
\end{align}

We observe that as $m \to \infty$, the fidelity goes to $1/q$. In this limit, diagramatically, the circuit is equivalent to a local circuit with 3 qudits. We discuss this family of circuits in the next subsection. As we shall see, the result we get there is compatible with the result here.

\subsection{Local Circuits} \label{sec:local_circuits}

In this Section, we consider the local circuits. A local circuit with $m$-columns and $n$ qudits are shown in Fig.~\ref{fig:local_circuit}. Without loss of generality, we choose $m$ and $n$ both to be even numbers.

Let us first consider the $m\leq n - 2$ case. The lattice diagram corresponding to the case of $n \geq m+2$ is given by
\begin{align}
    \adjincludegraphics[width=7cm,valign=c]{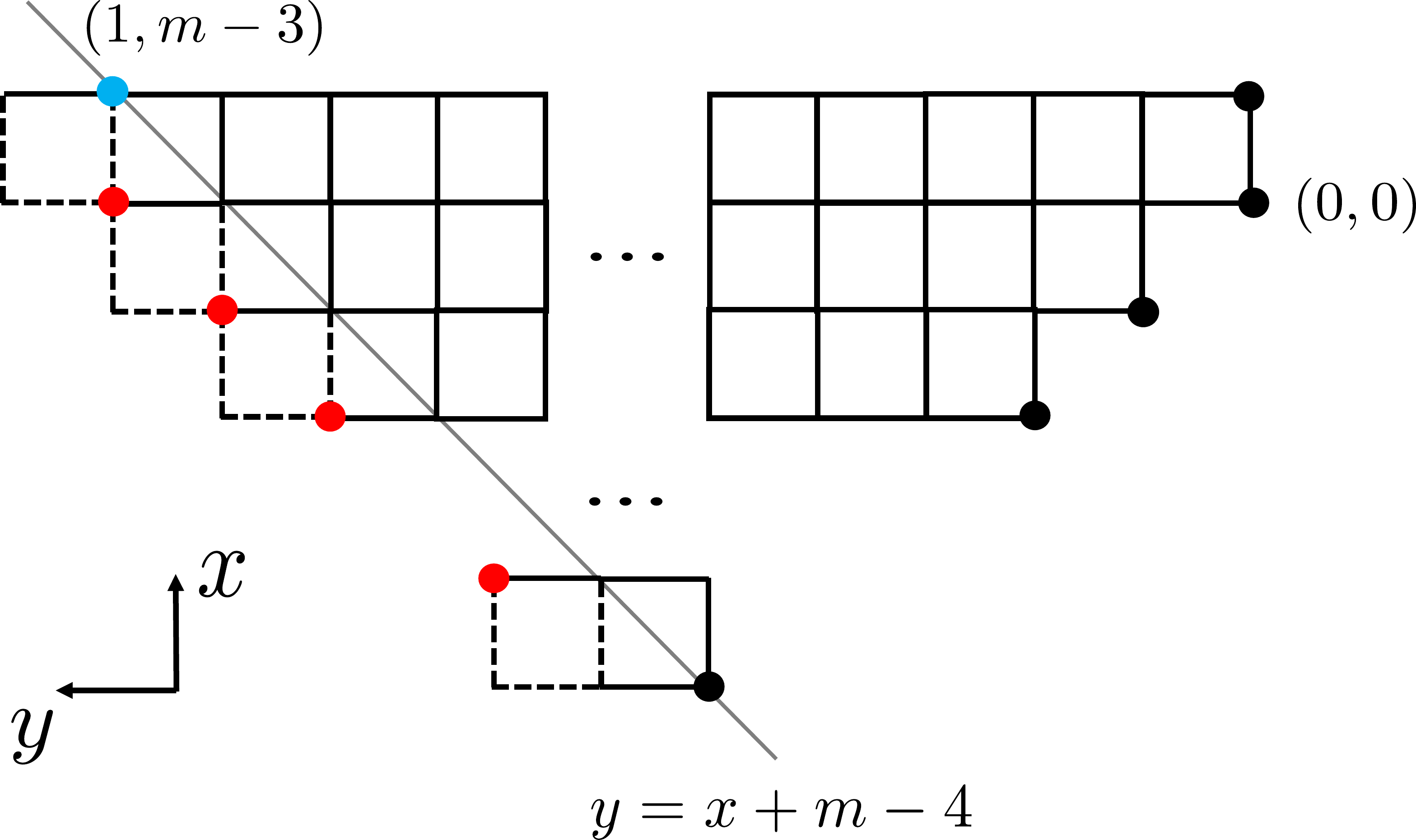} \notag
\end{align}

By applying Eq.~(\ref{Wein_3}), Eq.~(\ref{Wein_4}), Eq.~(\ref{general_formula}) and Eq.~(\ref{theorem_2}), we can obtain the fidelity:{\footnotesize
\begin{widetext}
\begin{align}
    F_1^m &= \left(\frac{q}{1+q^2}\right)^{m-2}\sum_{k=0}^{(m-4)/2} q^{m-4-2k} \left(  L_{1\ (-k,k)}^{(1,m-3)} + \sum_{l=1}^{k+1}\left(\frac{1+q^2}{q^2}\right)^l \sum_{\mathcal{I}_l}\left( L_{2\ {-k,k}}^{(i_1, i_1 + m - 4)} \prod_{j=1}^{l} L_{2\ (i_j+1, i_j+m-3)}^{(i_{j+1}, i_{j+1}+m-4)}\right)\right)\notag\\
    &+\left(\frac{q^2}{1+q^2}\right)^{m-2}\label{local_circuit_fidelity}
\end{align}
\end{widetext}}
Here we define $L_{1\ (-k,k)}^{(1,m-3)}$ to be $| L ((-k,k) \to (1, m-3)) | y \leq x + m - 5 $, and $L_{2\ {-k,k}}^{(i_1, i_1 + m - 4)}$ to be $| L ((-k,k) \to (i_1, i_1 + m - 4)) |  y \leq x + m - 4 |$. The last two terms are the paths that go through the line and the paths that do not go through the line. When $n \geq m + 2$ and $i \leq n - m$, we have the following identity
\begin{align}
    F_1^m =1
\end{align}
This means that the state of the qudit is perfectly restored after applying the rewinding protocol.

Now consider the $m \geq n$ case. The corresponding lattice diagram is given by
\begin{align}
    \adjincludegraphics[width=7cm,valign=c]{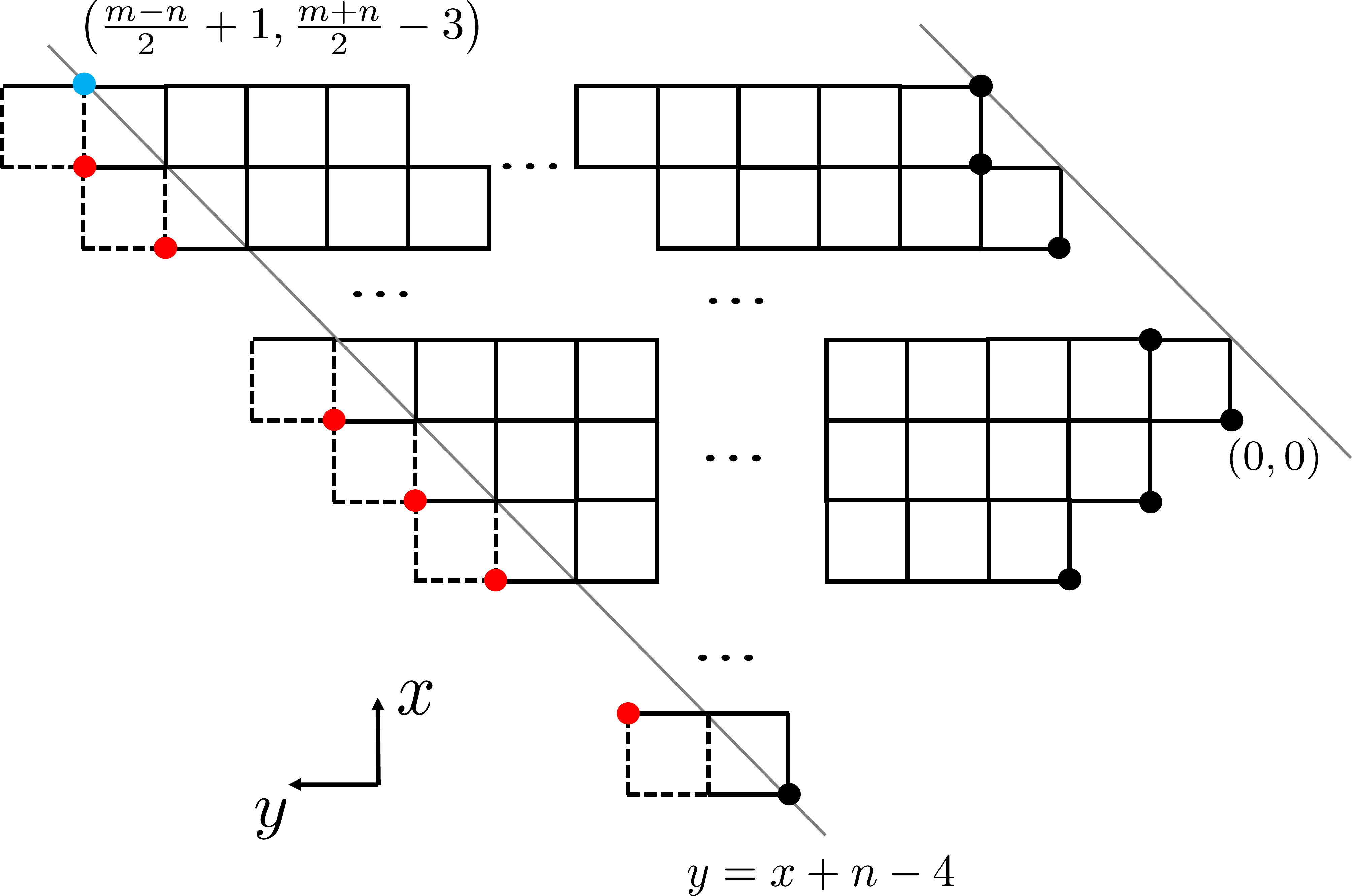}. \notag
\end{align}
By applying Eq.~(\ref{Wein_3}), Eq.~(\ref{Wein_4}), Eq.~(\ref{general_formula}) and Eq.~(\ref{theorem_2}), the expression for the fidelity can be derived:
\begin{widetext}
\begin{align}
    F_1^m &= \frac{1}{q} \left(\frac{q^2}{1+q^2}\right)^{n-2} + \sum_{k=1}^{(m-n)/2}\left(\frac{q}{1+q^2}\right)^{m-2k}q^{n-3} \left( L_{1\ (k,k)}^{((m-n)/2+1,(m+n)/2-3)} \vphantom{\left( L_{2\ (k,k)}^{i_1,i_1+n-4} \prod_{j=1}^{l} L_{(2\ i_j+1,i_j+n-3)}^{(i_{j+1},i_{j+1}+n-4)}\right)} \right. \notag \\
    &\left.+ \sum_{l=1}^{(m-n+2-2k)/2} \left(\frac{1+q^2}{q^2}\right)^l \sum_{\mathcal{I}_l} \left( L_{2\ (k,k)}^{i_1,i_1+n-4} \prod_{j=1}^{l} L_{(2\ i_j+1,i_j+n-3)}^{(i_{j+1},i_{j+1}+n-4)}\right)\right) \notag \\
    &+ \sum_{k=0}^{(n-4)/2} \left(\frac{q}{1+q^2}\right)^{m-2} q^{n-4-2k} \left( L_{1\ (-k,k)}^{((m-n)/2+1,(m+n)/2-3)} \vphantom{\left( L_{2\ (k,k)}^{i_1,i_1+n-4} \prod_{j=1}^{l} L_{(2\ i_j+1,i_j+n-3)}^{(i_{j+1},i_{j+1}+n-4)}\right)} \right. \notag \\
    &\left.+ \sum_{l=1}^{(m-n+2k+2)/2} \left(\frac{1+q^2}{q^2}\right)^l \sum_{\mathcal{I}_l} \left( L_{2\ (-k,k)}^{(i_1,i_1+n-4)} \prod_{j=1}^{l} L_{2\ (i_j+1,i_j+n-3)}^{(i_{j+1},i_{j+1}+n-4)}\right)\right)
\end{align}
\end{widetext}
Here we define $L_{1\ (k,k)}^{((m-n)/2+1, (m+n)/2-3)}$ to be $| L ((k,k) \to ((m-n)/2+1, (m+n)/2-3)) | x + n - 5 \geq y \geq x - 1 |$ and $L_{2\ (-k,k)}^{(i_1,i_1+n-4)}$ to be $| L ((-k,k) \to (i_1,i_1+n-4)) | x + n - 4 \geq y \geq x - 1 |$. The numerical results are shown in FIG.~\ref{fig:hybrid_numerical} (d). We notice that as $m \to \infty$, the fidelity approaches $1/q$, as expected.

\section{Recycling with imperfect gates}
\label{sec:recycling_imperfect}

In this Section, we explain how our calculations can be modified in the presence of noise. We first briefly explain how the calculations in Section~\ref{sec:diagrammatics} change, which is applicable to arbitrary quantum circuits. We will then focus on the calculation in the convolutional limit, which remains analytically tractable. 

Without loss of generality, we consider an error model in which the unitary $U_i$ is followed by a quantum channel acting on the qudits that $U_i$ acts on. To make the calculation tractable, we assume that this channel is a tensor product over the qudits that $U_i$ acts on. This channel can be conveniently represented by its Kraus representation. Specifically, this noise channel can be written as
\begin{align}
    \Phi (\cdot) = \sum_{k} E_{k} (\cdot) E_{k}^{\dagger}.
\end{align}
We shall also make the following simplifying assumptions; the error model is a tensor product of a single-qudit error model, and the model itself is assumed to be the same for every gate. We shall denote the set of Kraus operators as $\{E_k\}$.


These Kraus operators change the inner products involving $\tau_1$, $\tau_2$ and $\sigma$ in Eq.~(\ref{wein_diag}). For the convolutional circuit we show in Eq.~(\ref{lattice_node}), we summarize the rules in the following\footnote{For hybrid circuits and local circuits, the rule is different and more complicated. More general rules of diagrams are discussed in the appendix \ref{appx:1}.}.
\begin{align}
 \adjincludegraphics[width=1.4 cm,valign=c]{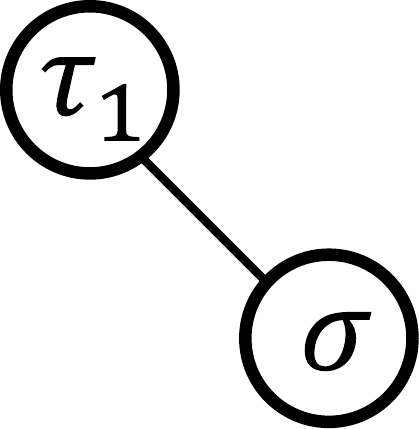} &= \begin{cases}
    q^2, \quad \tau_1 = \sigma = \mathbf{1} \\
    \alpha q^2, \quad \tau_1 = \sigma = \mathbf{s} \\
    q, \quad \text{otherwise}
    \end{cases}\\
    \adjincludegraphics[width=1.4 cm,valign=c]{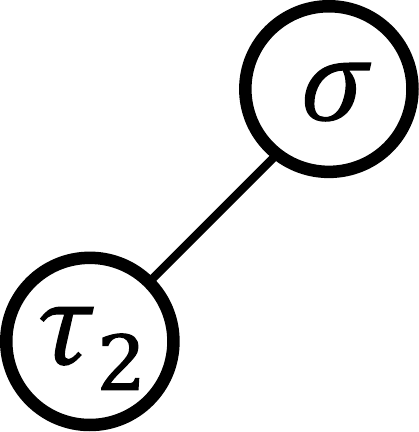} &= \begin{cases}
    q^2, \quad \tau_1 = \sigma = \mathbf{1} \\
    \beta q^2, \quad \tau_1 = \sigma = \mathbf{s} \\
    q, \quad \text{otherwise}
    \end{cases},
\end{align}
in which we define
\begin{align}
    \alpha = \frac{\sum_k |\mathrm{Tr}(E_{k})|^2}{q^2}, \quad \beta = \frac{\sum_{k,k'} |\mathrm{Tr}(E_{k}E_{k'})|^2}{q^2}.
\end{align}
We note that $\alpha$ is in fact the {\it entanglement fidelity} \cite{Schumacher1996}. Therefore, the diagrammatic rules in Eq.~(\ref{Wein_3}) become
\begin{align}
    \adjincludegraphics[width=1.1cm,valign=c]{diag_3.pdf} = 
    \begin{cases}
    1, \ (\tau_1, \tau_2, \tau_3) = (\mathbf{1},\mathbf{1},\mathbf{1}) \\
    0, \ (\tau_1, \tau_2, \tau_3) = (\mathbf{1},\mathbf{1},\mathbf{s}) \\
    \frac{q(q^2 - \beta)}{q^4-1}, \ (\tau_1, \tau_2, \tau_3) = (\mathbf{1},\mathbf{s},\mathbf{1}) \\
     \frac{q(\beta q^2 - 1)}{q^4-1}, \ (\tau_1, \tau_2, \tau_3) = (\mathbf{1},\mathbf{s},\mathbf{s}) \\
      \frac{q(q^2 - \alpha)}{q^4-1}, \ (\tau_1, \tau_2, \tau_3) = (\mathbf{s},\mathbf{1},\mathbf{1}) \\
       \frac{q(\alpha q^2 - 1)}{q^4-1}, \ (\tau_1, \tau_2, \tau_3) = (\mathbf{s},\mathbf{1},\mathbf{s}) \\
        \frac{q^2(1 - \alpha \beta)}{q^4-1}, \ (\tau_1, \tau_2, \tau_3) = (\mathbf{s},\mathbf{s},\mathbf{1}) \\
         \frac{\alpha \beta q^4 - 1}{q^4-1}, \ (\tau_1, \tau_2, \tau_3) = (\mathbf{s},\mathbf{s},\mathbf{s}) \\
    \end{cases} \label{eq:noiseless_diagram}
\end{align}

With this change, the simplifications we made in Section~\ref{sec:Mapping} becomes no longer valid. Now multiple domain walls are allowed configurations, and as such, simply counting the number of paths is not enough. In particular, isolated collection of $\mathbf{1}$s can now appear as islands surrounded by $\mathbf{s}$ nodes. 

\subsection{Convolutional circuit}

Fortunately, in the convolutional circuit, the calculation of the averaged fidelity nevertheless remains tractable. Note that Eq.~(\ref{lattice_node}) can be effectively regarded as a one-dimensional chain and each trivalent of the diagram can be represented by a $2 \times 2$ transfer matrix. Taking into account of the boundary conditions, we can define the following transfer matrix.
\begin{align}
    T &= \adjincludegraphics[width=2.5cm,valign=c]{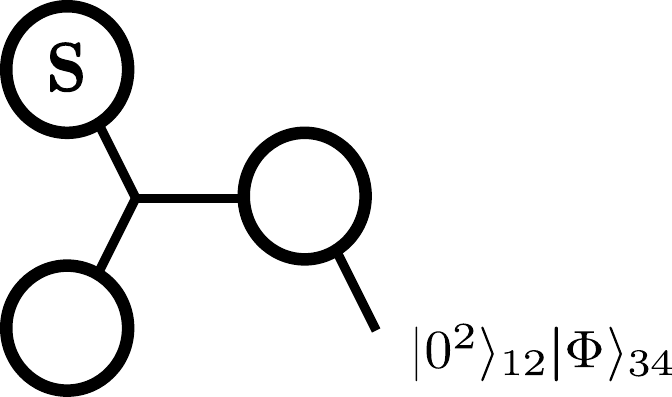} \notag \\
    &= \begin{pmatrix}
\mathbf{1} \to \mathbf{1} & \mathbf{s} \to \mathbf{1}\\
\mathbf{1} \to \mathbf{s} & \mathbf{s} \to \mathbf{s} \end{pmatrix} \notag \\ &=\begin{pmatrix}
\frac{q^2(q^2 - \alpha)}{q^4-1} & \frac{(1-\alpha \beta)q^3}{q^4-1}\\
\frac{q(\alpha q^2-1)}{q^4-1} & \frac{\alpha \beta q^4 - 1}{q^4 - 1}
\end{pmatrix}
\end{align}

We can now diagonalize the transfer matrix
\begin{align}
    T = P D P^{-1},
\end{align}
where
\begin{align}
   D = \begin{pmatrix}
    1 & 0\\
    0 & \frac{\alpha q^2 (\beta q^2 - 1)}{q^4 - 1}
    \end{pmatrix},\
    P = \begin{pmatrix}
    \frac{(1 - \alpha \beta)q^3}{\alpha q^2 - 1} & -\frac{1}{q}\\
    1 & 1
    \end{pmatrix}
\end{align}

Using this decomposition, we can now compute the averaged fidelity of the first qudit.  Define the $\mathbf{1}$ state to be the column vector $(1,0)$ and the $\mathbf{s}$ state to be the vector $(0,1)$. The averaged fidelity is then
\begin{align}
    \mathcal{F}_1 = A + B,
\end{align}
where
\begin{align}
    \begin{pmatrix}
    A\\
    B
    \end{pmatrix} = \frac{1}{q} P D^{n-k} P^{-1} \begin{pmatrix}
    1\\
    0
    \end{pmatrix}.
\end{align}
Plugging in the expressions for $P$ and $D$, we obtain
{\small \begin{align}
    \mathcal{F}_1 = \frac{(1 - \alpha \beta)q^3 + (\alpha q^2 - 1) \left(1 - \frac{q-1}{q} \left(\frac{\alpha q^2 (\beta q^2 - 1)}{q^4 - 1}\right)^{n-2}\right)}{(1 - \alpha \beta)q^4 + \alpha q^2 - 1}.
\end{align}}
When $n$ is very large, this converges to the following expression.
\begin{align}
    \sup \mathcal{F}_1 = \frac{(1-\alpha \beta)q^3 + \alpha q^2 -1}{(1-\alpha \beta)q^4 + \alpha q^2 -1}
\end{align}

This calculation can be straightforwardly generalized to the $i$'th qudit, yielding
\begin{widetext}
\begin{align}
    \mathcal{F}_i = \frac{(1 - \alpha \beta)q^3 + \alpha q^2 - 1}{(1 - \alpha \beta)q^4 + \alpha q^2 - 1} - \frac{\alpha q(\alpha q^2 - 1)(\beta q^2 - 1)}{(1+q)(1+q^2)[(1 - \alpha \beta)q^4 + \alpha q^2 - 1]}\left(\frac{\alpha q^2(\beta q^2 - 1)}{q^4 - 1}\right)^{n-i}
\end{align}
\end{widetext}

\subsection{Correlation}
We can also compute the correlation between two different reset qudits. This can be achieved by inserting the following modified transfer matrix at two locations $i$ and $j$:
\begin{align}
    T_0 &= \adjincludegraphics[width=1.8 cm,valign=c]{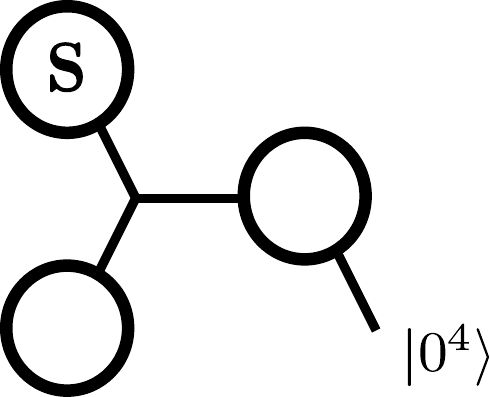} = \begin{pmatrix}
\frac{q(q^2 - \alpha)}{q^4-1} & \frac{(1-\alpha \beta)q^2}{q^4-1}\\
\frac{q(\alpha q^2-1)}{q^4-1} & \frac{\alpha \beta q^4 - 1}{q^4 - 1}
\end{pmatrix}
\end{align}
which can be diagonalized as $T_0 = P_0 D_0 P_0^{-1}$.

The fidelity of recycling the $i$-th and the $j$-th qudits can be formally written down as
\begin{align}
    \mathcal{F}_{ij} = q A + B,
\end{align}
where $A$ and $B$ are given by
\begin{align}
    \begin{pmatrix}
    A\\
    B
    \end{pmatrix} = \frac{1}{q} T^{j-2} T_0 T^{i-j-1} T_0 T^{n-i} \begin{pmatrix}
    1\\
    0
    \end{pmatrix}
\end{align}

The correlation between the $i$-th and the $j$-th qudit can be quantified in terms of connected correlation function
\begin{align}
    \mathcal{F}_{ij}^c = \mathcal{F}_{ij} - \mathcal{F}_i \mathcal{F}_j.
\end{align}
While the general expression for this correlation function is complicated, it simplifies in the $n\to \infty$ limit, assuming $\beta=1$. In this case, we have the expression
\begin{align}
    \lim_{\substack{n\to \infty\\ \beta \to 1}}\mathcal{F}_{ij}^c = \frac{(1 - \alpha )q^2(\alpha q^2 - 1)}{(q^2+1)((1-\alpha)q^2+1)^2}\left(\frac{\alpha q^2}{q^2+1}\right)^{i-j}
\end{align}
We see that the correlation function decays exponentially as the distance between two qudits increases. When we set $\alpha = \beta = 1$, the expressions of the noisy case becomes the expressions of the noiseless case in Eq.~\eqref{eq:noiseless_correlation}, as expected.

\section{Conclusion}
\label{sec:conclusion}

In this paper, we proposed a method to calculate the fidelity of the rewinding protocol in various random quantum circuits analytically. We established a connection between this problem and the counting of directed paths on graphs, which are determined by the shape of the circuits. 

We showed that in the convolutional limit, the fidelity approaches $1$ asymptotically and the rate of convergence is determined by a constant $\frac{q^2}{q^2+1}$. In the random quantum circuit limit, if the depth is smaller than the number of qudits, the protocol yields a unit fidelity. However, in the limit the circuit depth goes to infinity, the fidelity reduces to $1/q$.

We also derived two extra results in the convolutional limit, which may be of an independent interest. First, we derived the exact expressions for the correlations between recycled qudits and showed that it decays exponentially in the distance. Second, we derived the expressions in the presence of noise. In the presence of noise, the asymptotic value of the fidelity is no longer $1$, but becomes a number that depends on the details of the error model. Nevertheless, it approaches $1$ in the limit the noise vanishes, as expected. 

We remark that a major simplification of our analysis came from a connection between the problem at hand and the path counting problem. However, this correspondence appears to break down in the presence of noise. As we show in Appendix~\ref{appx:1}, the expression for the diagrammatics change. In particular, with this change, there can be multiple domain walls that can contribute to the fidelity. Physically, we expect the contributions from multi-domain wall contribution to be significantly smaller than the dominant contribution discussed in this paper, in the low error rate regime. As such, a natural question is to calculate this extra contribution perturbatively. We leave such analysis for future work.


\emph{Acknowledgements.} We thank Steve Flammia and Patrick Hayden for helpful discussions.

\bibliography{main}

\appendix 
\section{Detailed calculation} \label{calculation}

Consider a quantum wire that connects $d$-dimensional quantum gates. We have the following diagrammatic representation
\begin{align}
   \adjincludegraphics[width = 4 cm,valign=c]{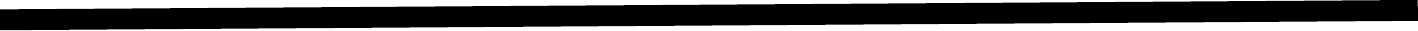} = \sum_{i=0}^{d-1} |i\rangle \langle i |.
\end{align}
It can be folded into two directions,
\begin{align}
   \adjincludegraphics[width = 1.5 cm,valign=c]{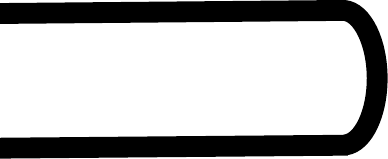} = \sum_{i=0}^{d-1} |ii\rangle, \quad \adjincludegraphics[width = 1.5 cm,valign=c]{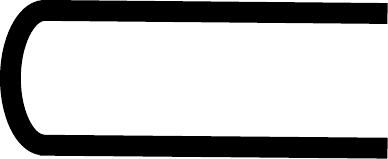} = \sum_{i=0}^{d-1} \langle ii|
\end{align}
Now, consider a $d$-dimensional quantum gate $U_{ij}$, where $i, j \in \{0, 1, ..., d-1\}$.
\begin{align}
   U_{ij} = \adjincludegraphics[width = 3 cm,valign=c]{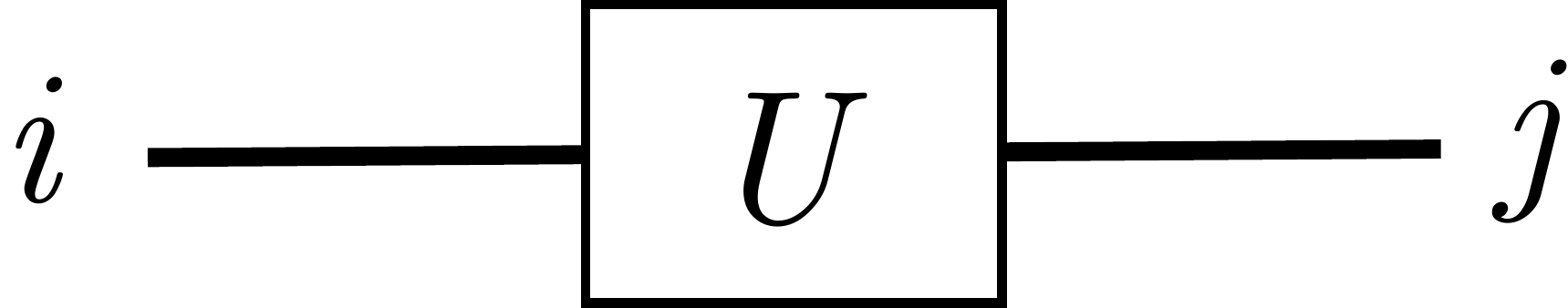}.
\end{align}
Therefore, Eq.~\eqref{eq:weingarten1} can be interpreted in the following way,
\begin{align}
     \int \mathrm{d}U\ \adjincludegraphics[width = 3 cm,valign=c]{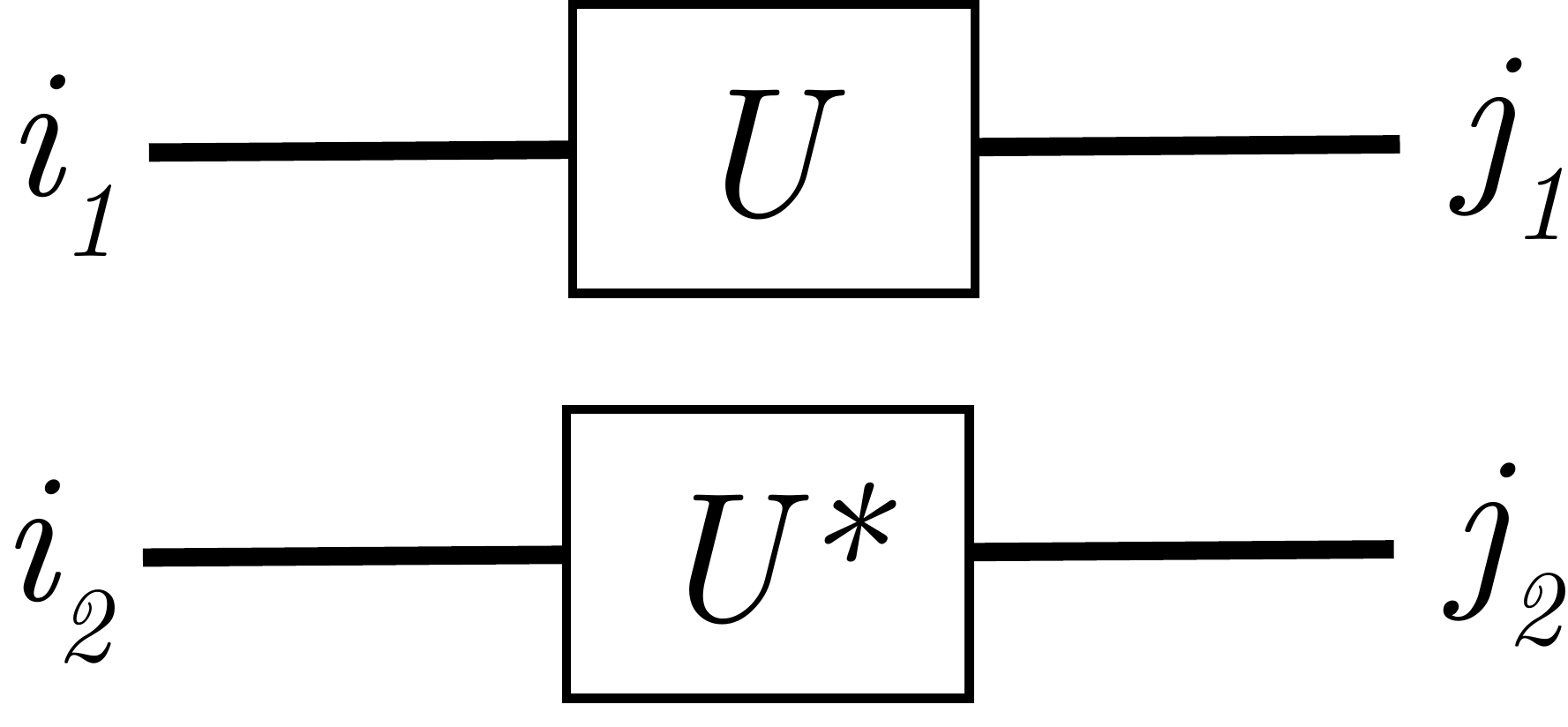} &= \frac{1}{d} \quad \adjincludegraphics[width = 2 cm,valign=c]{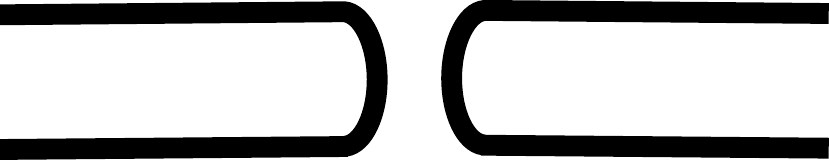} \notag\\&= \frac{1}{d} \sum_{i,j = 0}^{d-1} |ii\rangle \langle jj|
\end{align}
Similarly, Eq.~\eqref{eq:weingarten2} can be expressed equivalently as the following equation.

\begin{widetext}
\begin{align}
    \int \mathrm{d}U\ \adjincludegraphics[width = 3 cm,valign=c]{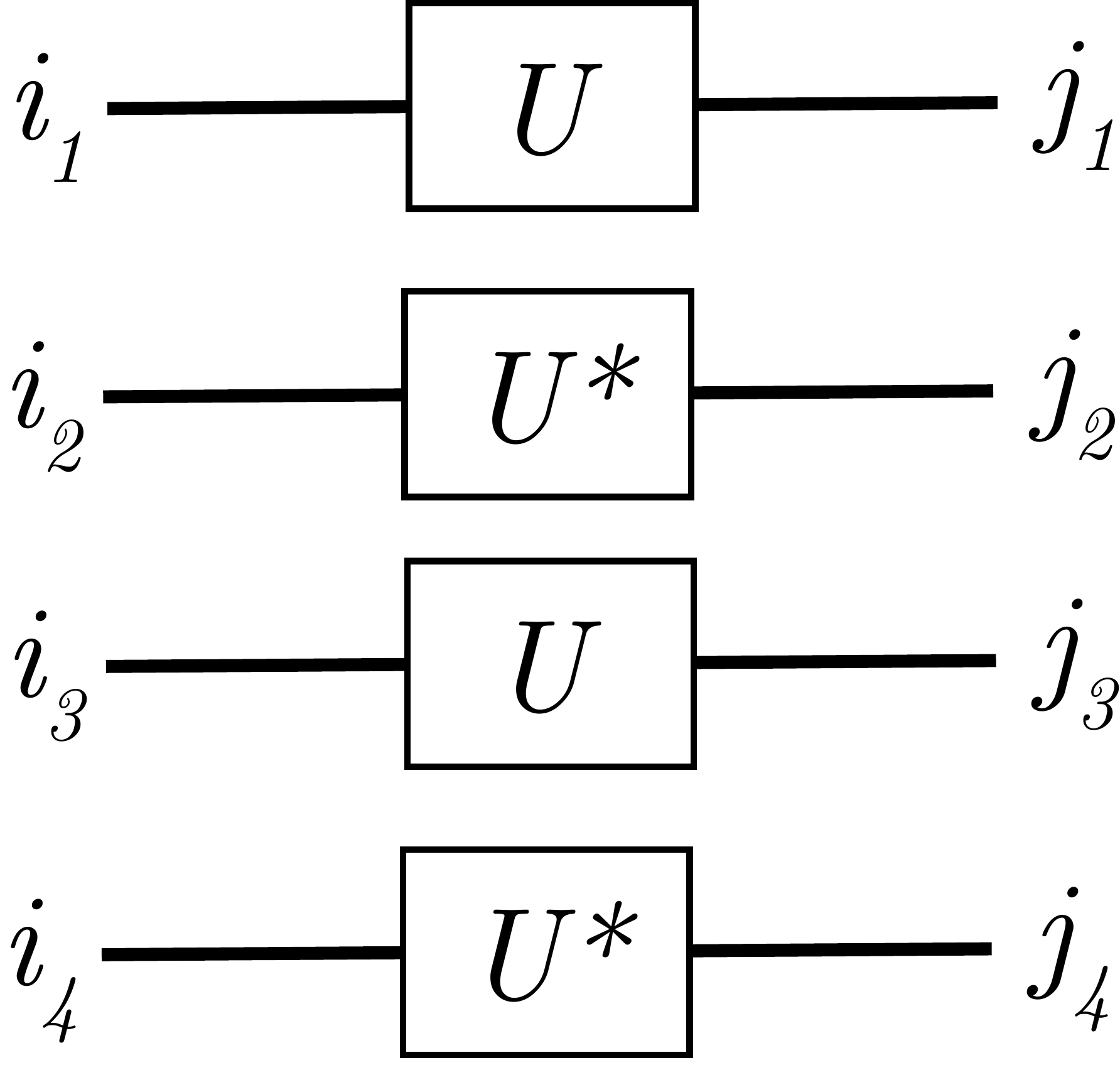} &= Wg(\mathbf{1}, d)\ \adjincludegraphics[width = 2.5 cm,valign=c]{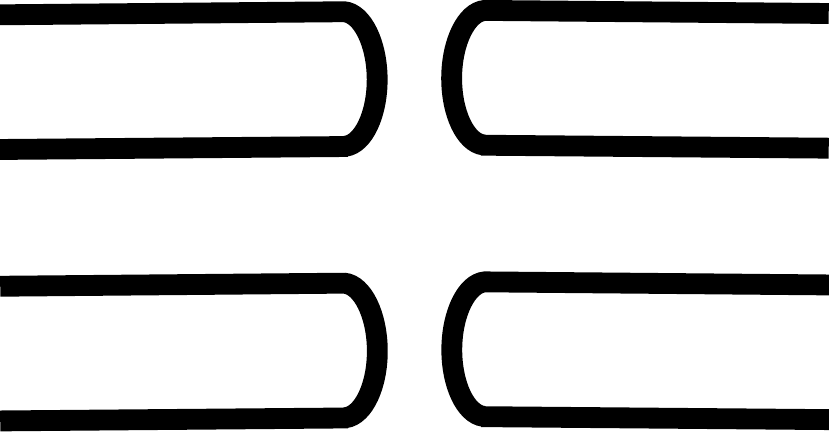} + Wg(\mathbf{s}, d)\ \adjincludegraphics[width = 2.5 cm,valign=c]{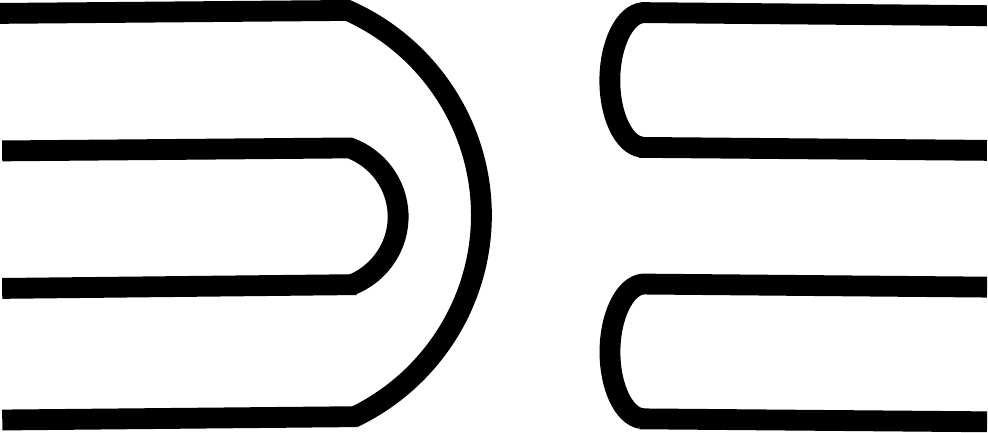} \notag \\
    &+ Wg(\mathbf{s}, d)\ \adjincludegraphics[width = 2.5 cm,valign=c]{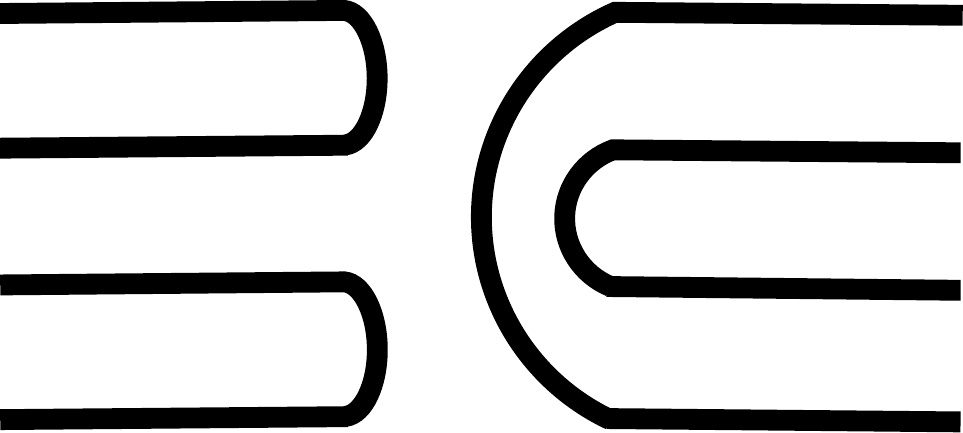} + Wg(\mathbf{1}, d)\ \adjincludegraphics[width = 2.5 cm,valign=c]{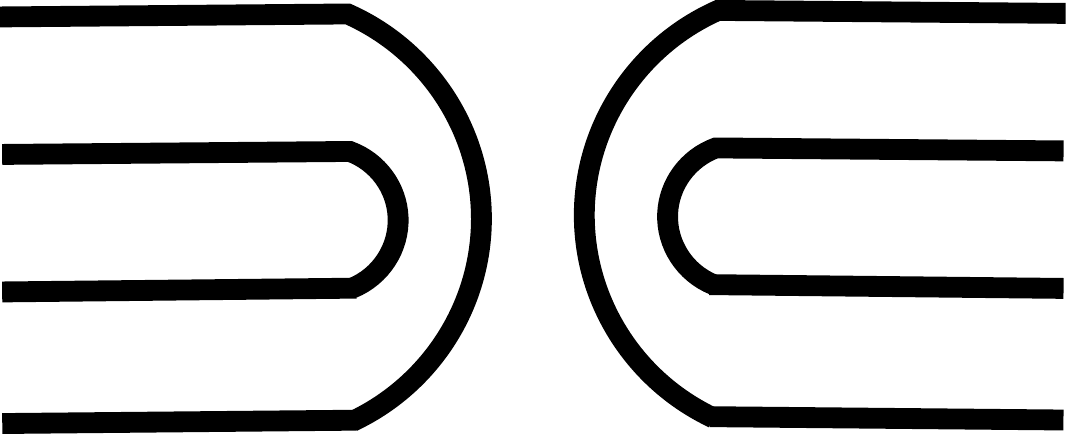} \label{eq:5}
\end{align}
\end{widetext}

It is important to note that, following the rule of Weingarten calculus, the indices of $U$ must be paired with the indices of $U^*$. As a result, there are four distinct types of pairings when dealing with four unitaries. 

The alignment of $U$ and $U^*$ follows Eq.~\eqref{eq:folding}. Since the $S_2$ group has only two elements, when the connection patterns on the left and right are the same, $\sigma \tau^{-1} = \mathbf{1}$. Conversely, when they are different, $\sigma \tau^{-1} = \mathbf{s}$. Recalling that under Eq.~\eqref{Wein_1}, we define $|\mathbf{1}\rangle = |\Phi\rangle_{12}|\Phi\rangle_{34}$ and $|\mathbf{s}\rangle = |\Phi\rangle_{14}|\Phi\rangle_{23}$, where $|\Phi\rangle_{ij} = \sum_{k=1}^{q} | k \rangle_{i}| k \rangle_{j}$, we can interpret Eq.~\eqref{eq:5} in terms of these definitions. This provides further insight into how Eq.~\eqref{Wein_1} follows.

\section{Diagrammatics: General case} \label{appx:1}

Here we provide the general form of the diagrams used in Section~\ref{sec:diagrammatics} in the presence of noise. Using the prescription in the main text, which involves folding the circuit diagram and averaging over the Haar measure, we obtain
\begin{align}
    \adjincludegraphics[width=1.0cm,valign=c]{diag_3.pdf} = 
    \begin{cases}
    1, \ (\tau_1, \tau_2, \tau_3) = (\mathbf{1},\mathbf{1},\mathbf{1}), \\
    0, \ (\tau_1, \tau_2, \tau_3) = (\mathbf{1},\mathbf{1},\mathbf{s}), \\
    \frac{q(q^2 - \beta_u)}{q^4-1}, \ (\tau_1, \tau_2, \tau_3) = (\mathbf{1},\mathbf{s},\mathbf{1}), \\
     \frac{q(\beta_u q^2 - 1)}{q^4-1}, \ (\tau_1, \tau_2, \tau_3) = (\mathbf{1},\mathbf{s},\mathbf{s}), \\
      \frac{q(q^2 - \beta_d)}{q^4-1}, \ (\tau_1, \tau_2, \tau_3) = (\mathbf{s},\mathbf{1},\mathbf{1}), \\
       \frac{q(\beta_d q^2 - 1)}{q^4-1}, \ (\tau_1, \tau_2, \tau_3) = (\mathbf{s},\mathbf{1},\mathbf{s}), \\
        \frac{q^2(1 - \beta)}{q^4-1}, \ (\tau_1, \tau_2, \tau_3) = (\mathbf{s},\mathbf{s},\mathbf{1}), \\
         \frac{\beta q^4 - 1}{q^4-1}, \ (\tau_1, \tau_2, \tau_3) = (\mathbf{s},\mathbf{s},\mathbf{s}). \\
    \end{cases}
\end{align}
Here the constants are defined as follows.
\begin{align*}
    \beta &= \frac{\sum_{k,k'}|\mathrm{Tr}(E_k E_{k'})|^2}{q^4} \\
    \beta_u &= \frac{1}{q^3} \sum_{k,k'} \langle 1 |_{u} \langle s |_d E_{k}^{\dagger} \otimes E_{k}^{T} \otimes E_{k'} \otimes E_{k'}^{*} |s\rangle_u |s\rangle_d \\
    \beta_d &= \frac{1}{q^3} \sum_{k,k'} \langle s |_{u} \langle 1 |_d E_{k}^{\dagger} \otimes E_{k}^{T} \otimes E_{k'} \otimes E_{k'}^{*} |s\rangle_u |s\rangle_d.
\end{align*}

We note that, unlike in Section~\ref{sec:recycling_imperfect}, we made no assumptions about the Kraus operators. That is, they are allowed to be arbitrary operators acting on two qubits, subject to the usual constraint that they define a channel.

\end{document}